\documentclass[showpacs,prd,preprintnumbers,nofootinbib,superscriptaddress
]{revtex4-2}

\usepackage{amsmath,amsthm,amssymb}
\usepackage{mathrsfs} 
\usepackage{upgreek}
\usepackage[normalem]{ulem} 
\usepackage{graphicx}
\usepackage{latexsym}
\usepackage{url,hyperref}
\usepackage{textcomp}
\usepackage[dvipsnames]{xcolor} 
\usepackage{slashed}
\usepackage{epstopdf}
\usepackage{placeins}
\usepackage{array}
\usepackage{cancel}
\usepackage{verbatim}
\usepackage{caption}
\usepackage{subcaption}
\usepackage[splitrule,hang,flushmargin]{footmisc} 
\renewcommand{\footnoterule}{%
  \kern -3pt
  \hrule width \textwidth height 1pt
  \kern 2pt
}

\usepackage{chngcntr}
\usepackage{pdfpages}

\usepackage{setspace} 
\AtBeginDocument{\setstretch{1.125}}

\usepackage{outlines} 
\usepackage{tensor} 
\usepackage{csquotes} 

\usepackage{blindtext}
\usepackage{tcolorbox}

\usepackage{soul}
\hypersetup{colorlinks=true,linkcolor=blue,citecolor=magenta,filecolor=magenta,urlcolor=blue}

\numberwithin{equation}{section}

\DeclareUnicodeCharacter{2212}{-} 

\usepackage{etoolbox}

\makeatletter
\patchcmd{\@outputpage@head}{\@ifx{\LS@rot\@undefined}{}{\LS@rot}}{}{}{}
\patchcmd{\frontmatter@abstract@produce} 
  {\vskip200\p@\@plus1fil
   \penalty-200\relax
   \vskip-200\p@\@plus-1fil}
  {}
  {}
  {}
\makeatother

\newcolumntype{C}{>{$}c<{$}} 
\newcolumntype{L}{>{$}l<{$}} 

\begin{document}
\title{Greybody factors of Proca fields in a Schwarzschild spacetime: A supplemental analysis based on decoupled master equations related to the Frolov-Krtouš-Kubizňák-Santos separation}

\author{Supanat Bunjusuwan}
\email{Supanatb64@nu.ac.th}
\affiliation{The Institute for Fundamental Study, \\
Naresuan University, Phitsanulok 65000, Thailand.}

\author{Chun-Hung Chen}
\email{chun-hungc@nu.ac.th}
\affiliation{The Institute for Fundamental Study, \\
Naresuan University, Phitsanulok 65000, Thailand.}

\begin{abstract}
\par Greybody factors for Proca fields in Schwarzschild black hole spacetime are investigated. The radial equations are derived by separating the field equations using vector spherical harmonics and decoupling the even-parity sector through a set of transformations related to the Frolov-Krtouš-Kubizňák-Santos separation in the static limit. Semi-analytical methods, including a rigorous bound and the Wentzel-Kramers-Brillouin approximation, are used to compute the transmission probabilities. In addition to reproducing known results, two distinctive features are identified. In the even-parity vector mode, a low-mass regime is found where the transmission probability exceeds that of the massless case for a set of common energy and angular momentum parameters. In the even-parity scalar mode, the massless limit reproduces the result of massless scalar perturbations and corresponds to a pure gauge mode in Maxwell theory. In the same mode, the transmission probability in the massive case is systematically lower than that of a massive scalar field with the same parameters.
\end{abstract}


\date{\today}
\maketitle

\section{Introduction}
The linear perturbation of a massive vector field in spherically symmetric black hole spacetimes has attracted attention since the early $\mathrm{21^{st}}$ century \cite{kono2006}, focusing on the stability and quasinormal modes of monopole modes and infinitely higher overtones in the Schwarzschild background. In 2012, Rosa and Dolan introduced a set of four vector spherical harmonics (VSH), which allowed the radial equations to be separated into one decoupled odd-parity equation and three coupled even-parity equations \cite{DolanRosa2012}. The quasinormal modes and quasi-bound states were studied, and the spectrum was generally classified into two vector-type polarizations and one scalar-type polarization. It was confirmed that the scalar-type polarization corresponds to the even-parity modes, while the vector-type polarizations include both even-parity and odd-parity modes.

Studies of the greybody factor and Hawking radiation for Proca fields in spherically symmetric black hole spacetimes of general dimensions were presented around the same time \cite{HerSamWan2012, HerSamWan2013}. An extension of the Proca perturbation to slowly rotating Kerr black holes was given in \cite{PaniCarGua2012}. Research on Proca perturbations in black hole spacetimes flourished; however, a common challenge in these studies lies in the difficulty of decoupling and separating the Proca equations in both spherically symmetric and Kerr-like black hole spacetimes. Consequently, most analyses rely on numerical strategies or impose specific conditions to deal with the coupled or non-separable equations.

The problem of separation and decoupling for the Proca field persisted for decades, until Lunin proposed an ansatz for the separation variable in the general dimensional Myers-Perry-(A)dS geometry in 2017 \cite{Lunin2017}. In the next year, Frolov-Krtou$\mathrm{\check{s}}$-Kubiz$\mathrm{\check{n}\acute{a}}$k-Santos (FKKS) further generalized the ansatz for the separability of massive vector fields in the Kerr-NUT-(A)dS black hole spacetime \cite{FKKSPRL2018, FKK2018, KRTOUS20187}. Their works not only established separability for massless and massive vector fields in general-dimensional Kerr-like black holes but also yielded decoupled master equations in the static limit for spherically symmetric black holes. The instability of Proca perturbations was studied shortly after the FKKS separation in \cite{DolanInsta2018}, and transformations for decoupling the even-parity equations in the static limit were developed in \cite{PreDol2020, FerHilLeoCar2022}. These studies primarily focused on quasinormal modes and quasi-bound states, rather than greybody factors, although all these phenomena share the same master equations but differ in boundary conditions. The lack of investigations into greybody factors of Proca perturbations in spherically symmetric black hole backgrounds following the FKKS separation forms the main motivation of this work. While numerical results for this problem were previously presented in \cite{HerSamWan2012}, they considered Proca masses starting from $\mu = 0.3$ and lacked results for smaller masses. This region of the parameter space is the focus of this article. Specifically, we demonstrate that the transmission probability for massive particles can exceed that of massless particles in this regime, which contrasts with the general expectation for greybody factors of fields with the same spin.   

Our strategy proceeds as follows. The master equations for Proca perturbations in Schwarzschild spacetime are obtained through VSH separation \cite{DolanRosa2012}, and further decoupled by applying the FKKS-based transformation in the static limit \cite{PreDol2020, FerHilLeoCar2022}. We reduce the resulting equations to Schrödinger-like form and study the greybody factor using the rigorous bound method \cite{BooVi2008} and the Wentzel-Kramers-Brillouin (WKB) approximation \cite{IyerWill1987, WillGuinn1988}. The motivation for computing both methods is to verify the accuracy and consistency of our results, particularly in the lower transmission probability region where the WKB method may exhibit inconsistencies \cite{BoChNgW2021}. The rigorous bound is based on a mathematical inequality that provides limits for quantum scattering in one-dimensional Schrödinger systems \cite{PhysRevA.59.427}, and was applied to black hole perturbation theory in \cite{BooVi2008}. It yields lower bounds on the greybody factor and upper bounds on the reflection probability, ensuring that the true value lies within a specific range. Further developments can be found in \cite{BoNgVi2014, Boonserm_2023}. The WKB approximation is a general analytical technique widely used in quantum mechanics and wave propagation problems. In the context of black hole perturbation theory, it provides a systematic framework for analyzing wave equations under different boundary conditions, such as those associated with quasinormal modes and transmission probabilities. The third-order WKB method in this context was developed in \cite{IyerWill1987} and was later extended to include higher-order corrections in \cite{KonoWKB2003, Kono2019WKB}. 

This article is organized as follows. In Ch.~\ref{sec:ProcaSchwarz}, we present the derivation of Schrödinger-like radial equations for the Proca field in Schwarzschild spacetime. The behaviors of the effective potentials for vector-type and scalar-type polarizations are also summarized. In Ch.~\ref{sec:bound}, we derive bounds for the greybody factors and discuss the effectiveness of the “strict” and “less strict” bounds, and their relation to the effective potentials. In Ch.~\ref{sec:WKB}, we study greybody factors using the WKB approximation. Owing to the distinct mathematical forms of the different modes, separate treatments are required for the odd-parity, even-parity scalar, and even-parity vector modes, which are discussed in each subsection. In Ch.~\ref{sec:conclu}, we present the conclusions and discuss possible directions for future work. Throughout this paper, we adopt natural units by setting the speed of light, gravitational constant, and black hole mass to unity, $c=G=M=1$.

\section{Proca field in Schwarzschild black hole spacetime \label{sec:ProcaSchwarz}}
\subsection{Schr\"{o}dinger-like radial equation}
The metric element of a Schwarzschild black hole is given by
\begin{equation}\label{Schwarzs}
	ds^{2}=-fdt^{2}+f^{-1}dr^{2}+r^2(d{\theta}^2+\sin^2\theta d \phi^2), 
\end{equation}
where $f\equiv f(r)=1-2/r$. The Proca field equation is
\begin{equation} \label{field}
	\nabla_\alpha F^{\alpha\nu} =\mu^2 A^\nu,
\end{equation}
where $F_{\alpha\nu}=\nabla_{\alpha}A_{\nu}-\nabla_{\nu}A_{\alpha}$ is the antisymmetric field strength tensor, $A_{\nu}$ is the vector potential, $\mu$ is the mass parameter of the Proca field and $\nabla_{\alpha}$ is the covariant derivative. The imposing of the Lorenz condition $\nabla_{\nu}A^{\nu}=0$ can be determined by taking the covariant derivative $\nabla _\nu$ on both sides of Eq.~(\ref{field}) yields 
\begin{equation}
	\nabla_\nu \nabla_\alpha F^{ \alpha \nu}=\mu^2\nabla_\nu A^\nu.
\end{equation}
Using the antisymmetry of the field strength tensor, we obtain
\begin{eqnarray}
\nabla_\nu \nabla_\alpha F^{ [\alpha \nu]}=\frac{1}{2}[\nabla_\nu, \nabla_\alpha] F^{\alpha\nu} = R_{\nu\alpha}F^{\alpha\nu}=0.
\end{eqnarray}
Note that this condition applies generally to the Proca field in torsion-free spacetimes, not only in Ricci-flat backgrounds. It follows from the symmetry of the Ricci tensor and the antisymmetry of the field strength tensor, which ensures that the Lorenz condition is automatically satisfied. Substituting Eq.~(\ref{Schwarzs}) into Eq.~(\ref{field}), one obtains four partial differential equations corresponding to the free index as the coordinate basis $\nu\in\{t,r,\theta,\phi\}$. In general, the vector potential depends on all coordinates, $A_{\nu}=A_{\nu}(t,r,\theta,\phi)$. The separation using vector spherical harmonics (VSH) was introduced in \cite{DolanRosa2012}, defined as
\begin{equation}\label{VSH ansztz}
	A_\nu(t,r,\theta,\phi)=\frac{1}{r}\sum_{i=1}^{4}\sum_{lm} c_i\,u^{lm}_{(i)}(t,r)\,Z_\nu^{(i)lm}(\theta,\phi),
\end{equation}
with
\begin{eqnarray}
	Z_\nu ^{(1)lm}&=&[1,0,0,0]Y^{lm}(\theta,\phi),\label{VSH ansztz Z1}\\
	Z_\nu ^{(2)lm}&=&[0,f^{-1},0,0]Y^{lm}(\theta,\phi),\label{VSH ansztz Z2}\\
	Z_\nu ^{(3)lm}&=&\frac{r}{\sqrt{l(l+1)}}\left[0,0,\partial_\theta,\partial_\phi\right]Y^{lm}(\theta,\phi),\label{VSH ansztz Z3}\\ 
	Z_\nu ^{(4)lm}&=&\frac{r}{\sqrt{l(l+1)}} \left[0,0,\frac{1}{\sin\theta}\partial_\phi,\,-\sin\theta\partial_\theta\right]Y^{lm}(\theta,\phi),  \label{VSH ansztz Z4}
\end{eqnarray} 
where $c_1=c_2=1$, $c_3=c_4=(l(l+1))^{-1/2}$, and $Y^{lm}(\theta,\phi)$ is the scalar spherical harmonic satisfying the eigenvalue equation as
\begin{equation}\label{eigen}
	\left(\partial _\theta^2 + \cot \theta \,\partial_\theta + \frac{1}{\sin^2 \theta } \partial_\phi^2 \right)Y^{lm}(\theta,\phi)= -l(l+1)Y^{lm}(\theta,\phi).
\end{equation}
The separation of time dependence is given by $u^{lm}_{(i)}(t,r)=u_{(i)}e^{-i\omega t}$, where $\omega$ is the energy parameter when studying greybody factors and the remain $u_{(i)}$ are radial functions. The separated radial equations can be obtained as
\begin{eqnarray}
	&\hat{D}^{2} u_{(1)}& +  f'\left(-i \omega u_{(2)}-\partial_{r_{\star}} u_{(1)}\right) = 0,\label{Ra1} \\
	&\hat{D}^{2} u_{(2)}& + f'\left(-i \omega u_{(1)}-\partial_{r_{\star}}u_{(2)} \right)-\frac{2f^{2}}{r^{2}}\left( u_{(2)}-u_{(3)}\right) = 0, \label{Ra2}  \\
	&\hat{D}^{2} u_{(3)}&+\frac{2f}{r^2}l(l+1)u_{(2)} = 0, \label{Ra3} \\
	&\hat{D}^{2} u_{(4)}& = 0, \label{Ra4}
\end{eqnarray}
where primes denote $\frac{\partial}{\partial r}$, $\partial_{r_{\star}}\equiv f\frac{\partial}{\partial r}$ represents the partial derivative to the tortoise coordinate, and  
\begin{equation}
\hat{D}^{2}=\partial_{r_{\star}}^2+\omega^2-f\left(\frac{l(l+1)}{r^2}+\mu^2\right).
\end{equation}
The separated Lorenz condition reads
\begin{equation}\label{LorenzCondi}
i \omega u_{(1)}+\partial_{r_{\star}}u_{(2)}+\frac{f}{r}\left(u_{(2)}-u_{(3)}\right)=0,
\end{equation} 
and the Eqs.~(\ref{Ra1}) to (\ref{LorenzCondi}) be considered as a set of master equations describing the linear perturbations of Proca fields around the Schwarzschild background. Notably, Eq.~(\ref{Ra4}) is fully separated and decoupled. Since Eq.~(\ref{VSH ansztz Z4}) acquires a factor of $(-1)^{l+1}$ under a parity transformation, Eq.~(\ref{Ra4}) can be identified as describing the odd-parity, or magnetic, modes. On the other hand, the parity transformation yields the same factor, $(-1)^l$, for Eqs.~(\ref{VSH ansztz Z1}) to (\ref{VSH ansztz Z3}); therefore, Eqs.~(\ref{Ra1}) to (\ref{Ra3}) represent the coupled even-parity, or electric, modes.

To decouple the radial equations, a set of transformations on $u_{(i)}$ for the even-parity modes was provided by Percival and Dolan \cite{PreDol2020}, and by Fernandes et al. \cite{FerHilLeoCar2022}, derived after the FKKS separation, which is given by
\begin{eqnarray}
    u_{(1)}&=&-\frac{ifr(\nu r \partial_r+\omega /f )\bar{R}}{q_r},\label{u1}\\
    u_{(2)}&=&\frac{fr(\partial_r-\omega \nu  r /f)\bar{R}}{q_r},\label{u2}\\
    u_{(3)}&=&l(l+1) \bar{R},\label{u3}
\end{eqnarray}         
where $q_r = 1 + \nu^2 r^2$, and $\nu$ is a constant that arises in the separation process of the FKKS ansatz. In the static limit, $\nu$ can be determined by matching the separated angular equation of the Proca field in the Kerr–NUT–AdS spacetime with Eq.~(\ref{eigen}). Nevertheless, we do not elaborate on the full separation procedure of the FKKS ansatz in this article. Instead, we treat $\nu$ as an undetermined parameter to be derived in the subsequent analysis. With these transformations, a single radial equation for the even-parity modes can be obtained by directly substituting Eqs.~(\ref{u2}) and (\ref{u3}) into Eq.~(\ref{Ra3}), which yields
\begin{equation}\label{REven}
    \left[\partial^2_{r_*}+\frac{2f}{rq_r}\partial_{r_*}+\left(\omega^2-f\left(\frac{l(l+1)}{r^2}+\mu^2 \right)-\frac{2f\omega\nu}{q_r}\right)\right]\bar{R}=0.
\end{equation}   
To verify the consistency of Eqs.~(\ref{Ra1}) and (\ref{Ra2}) with Eq.~(\ref{REven}), one may first express $u_{(3)}$ in terms of $u_{(1)}$ and $u_{(2)}$ by substituting the Lorenz condition, Eq.~(\ref{LorenzCondi}), into Eq.~(\ref{Ra2}); in doing so, Eqs.~(\ref{Ra1}) and (\ref{Ra2}) become coupled equations involving only $u_{(1)}$ and $u_{(2)}$. By further substituting Eqs.~(\ref{u1}) and (\ref{u2}), and taking a linear combination of $\left(-1/i\omega r\right) \times$ Eq.~(\ref{Ra1}) and $\left(-\nu/\omega\right) \times$ Eq.~(\ref{Ra2}), one can eliminate the third-order derivative term and recover Eq.~(\ref{REven}), thereby confirming the consistency. Next, the constant $\nu$ can be determined by substituting Eqs.~(\ref{u1}) to (\ref{u3}) into the Lorenz condition, Eq.~(\ref{LorenzCondi}), and we obtain
\begin{equation}
\left[\partial_{r_{\star}}^{2}+\frac{2f}{rq_{r}}\partial_{r_{\star}}+\left(\omega^{2}-3f\omega \nu-\frac{f q_{r} }{r^2} l\left(l+1\right)+\frac{2f\omega \nu^{3}r^{2}}{q_{r}}\right)\right]\bar{R}=0.\label{LorenzCondiTr}
\end{equation}
By comparing Eq.~(\ref{LorenzCondiTr}) with Eq.~(\ref{REven}), and omitting the first- and second-order derivative terms, one obtains
\begin{equation}
l\left(l+1\right) \nu^{2}+\omega \nu -\mu^{2}=0.\label{nunu}
\end{equation}
Solving this quadratic equation, we have
\begin{equation}
   \nu =\frac{-\omega}{l(l+1)}\left(\frac{1\pm \sqrt{1+4l(l+1)\mu^2/\omega^2}}{2}\right), \label{nu}   
\end{equation}
which reproduces the result obtained by taking the static limit of the angular equation in the Kerr–NUT–AdS background and matching the spherical harmonics, as shown in \cite{DolanInsta2018}. Note that the $\pm$ sign refers to the ``vector" and ``scalar" modes, respectively, based on the corresponding effective potentials and the physical behavior in the massless limit, which will be more clearly presented in later sections. For the convenience of subsequent analysis, we further simplify the even-parity equation, Eq.~(\ref{REven}), into a Schrödinger-like form by setting $\bar{R}(r) = A(r) R(r)$, where 
\begin{equation}
       A(r)=\frac{\sqrt{q_{r}}}{r}.
\end{equation}
With this transformation applied to the even-parity equation, we may summarize its form accordingly. For the Proca field in Schwarzschild black hole spacetime, the radial equations for both even-parity and odd-parity modes include a Schrödinger-like form
\begin{equation}
    \left[\partial_{r_*}^2+\omega^2-V_{eff}^{(modes)}\right]R =0,\label{SLR}
\end{equation}
where $V_{eff}^{(modes)}$ denotes the effective potentials, which depend on $(r,l,\omega,\mu)$. The explicit expressions are
\begin{eqnarray}
V_{eff}^{(odd)}&=&f\left(\frac{l(l+1)}{r^2}+\mu^2\right),\label{VSLRO}\\
V_{eff}^{(even-scalar)}&=&f\left(\frac{l(l+1)}{r^2}+\mu^2\right) +\left(\partial_{r_*}\frac{f}{rq_{r}^{(-)}}\right)+\left(\frac{f}{rq_{r}^{(-)}}\right)^2+\frac{2f\omega\nu_{-}}{q_{r}^{(-)}},\label{VSLRES}\\
V_{eff}^{(even-vector)}&=&f\left(\frac{l(l+1)}{r^2}+\mu^2\right )+\left(\partial_{r_*}\frac{f}{rq_{r}^{(+)}}\right)+\left(\frac{f}{rq_{r}^{(+)}}\right)^2+\frac{2f\omega\nu_{+}}{q_{r}^{(+)}},\label{VSLREV}
\end{eqnarray}
where $\nu_{\pm}$ represents the $\pm$ sign in the bracket of Eq.~(\ref{nu}), and $q_{r}^{(\pm)} \equiv 1 + \nu_{\pm}^{2} r^{2}$. Since the radial equations have been simplified into a Schrödinger-like form, this allows us to study the behavior of the effective potentials and to apply both the rigorous bound and the WKB approximation for obtaining the greybody factors in the following sections.

\subsection{The effective potentials}\label{sec:effp}
\begin{figure}
    \centering
    \includegraphics[width=0.5\linewidth]{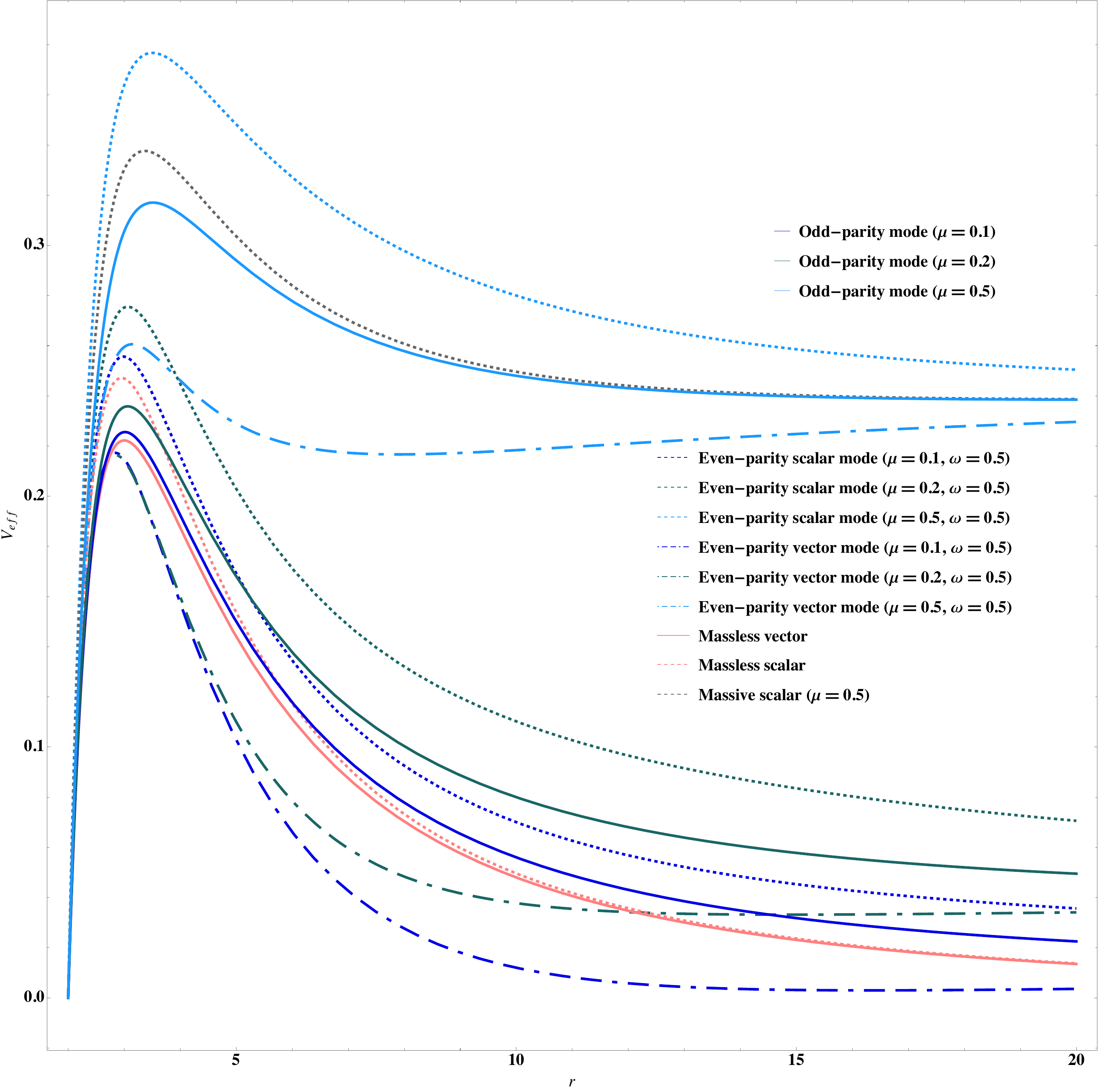}
    \caption{Comparison of the effective potentials for the three modes with fixed parameters $l=2$ and $\omega=0.5$.}
    \label{fig:effectivepotential}
\end{figure}

To compare the behavior across all modes, we begin with the odd-parity case. The effective potential in this mode is of the Regge–Wheeler type, and Eq.~(\ref{VSLRO}) always exhibits a maximum peak and a local minimum, with the corresponding radial positions given by
\begin{equation}\label{rmami}
r_{max}=\frac{l+l^2-\sqrt{l^4+2 l^3-12 \mu ^2 l^2+l^2-12 \mu ^2 l}}{2 \mu ^2}\ \ ; \ \ r_{min}=\frac{l+l^2+\sqrt{l^4+2 l^3-12 \mu ^2 l^2+l^2-12 \mu ^2 l}}{2 \mu ^2},
\end{equation}
 note that these expressions are valid only for cases with $\mu \neq 0$, and the condition for the existence of barrier-like effective potentials, i.e., $V_{eff}^{(\text{odd})}|_{\text{max}} > \mu^{2}$, can be obtained as $\mu < \sqrt{l(l+1)}/4$ for the odd-parity modes. In the case of $l = 1$, this condition yields a maximum Proca mass of $\mu \lesssim 0.353\ldots$, which corresponds to a relatively narrow parameter range for comparison with other modes. Therefore, we focus on the $l = 2$ case, where the barrier-like condition for the odd-parity mode becomes $\mu \lesssim 0.612\ldots$. In Fig.~\ref{fig:effectivepotential}, we fix $l = 2$, vary the Proca mass with $\mu = 0.1$, $0.2$, $0.5$, and present the effective potentials for the odd-parity mode as solid curves. For comparison, the effective potential of the massless spin-1 Regge–Wheeler equation is also shown as a pink solid curve. One may observe that, for the odd-parity modes, the maximum of the effective potential increases as the Proca mass $\mu$ increases, which is a typical behavior for massive perturbations in Schwarzschild black hole spacetimes.
 
\begin{figure}
    \centering
\includegraphics[width=0.5\linewidth]{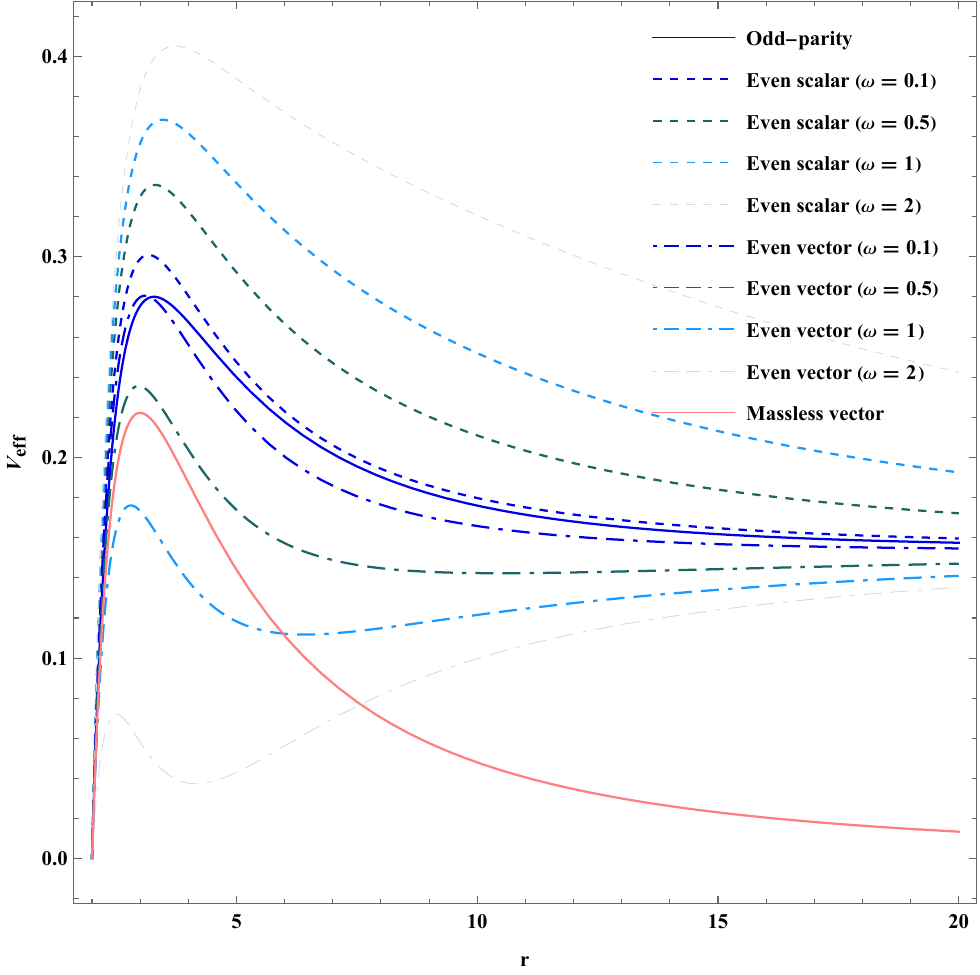}
    \caption{Comparison of the effective potentials for the three modes with fixed parameters $l=2$ and $\mu=0.4$.}
\label{fig:effectivepotentialfixlmu}
\end{figure}

For the even-parity modes, as shown in Eqs.~(\ref{VSLRES}) and (\ref{VSLREV}), the effective potentials are $\omega$-dependent, reflecting the energy dependence of the Proca particles. Energy-dependent effective potentials have also been observed in other perturbative contexts, such as massive Dirac perturbations \cite{Cho2003}, spin-3/2 perturbations \cite{PhysRevD.100.104018} in spherically symmetric black hole spacetimes, and the Teukolsky equation \cite{PhysRevLett.29.1114}. We illustrate the behavior of the $\omega$-dependent effective potentials by fixing $l = 2$ and $\omega = 0.5$, while varying $\mu = 0.1$, $0.2$, and $0.5$, as shown in Fig.~\ref{fig:effectivepotential}. Similarly, in Fig.~\ref{fig:effectivepotentialfixlmu}, we fix $l = 2$ and $\mu = 0.4$, and vary $\omega = 0.1$, $0.5$, $1$, and $2$. For ease of comparison, the even-parity scalar modes are shown with dotted lines, and the even-parity vector modes with dash-dotted lines. The general behavior of the even-parity modes is similar to that of the odd-parity modes, featuring a local maximum followed by a local minimum (a dip after the peak barrier), and asymptotic convergence to $\mu^{2}$ as $r \rightarrow \infty$. Due to the fact that taking first derivatives of Eqs.~(\ref{VSLRES}) and (\ref{VSLREV}) yields higher-order algebraic equations, general analytical expressions for the radial positions of the local extrema cannot be obtained explicitly. Nevertheless, the values of $r_{max}$ and $r_{min}$ can still be determined numerically by specifying the parameters $l$, $\mu$, and $\omega$.

More specifically, for the even-parity scalar modes, the series expansion of Eq.~(\ref{VSLRES}) in the limit $\mu\rightarrow 0$ is
\begin{equation}
V_{eff}^{(even-scalar)} \sim f\left( \frac{l(l+1)}{r^2}+\frac{2}{r^{3}}\right) + 3 f \mu^{2} + \mathcal{O}\left(\mu^{4}\right),\label{VSLRESep}
\end{equation}
where the leading-order approach to the effective potential of the Regge-Wheeler scalar equation. Therefore, the even-parity scalar modes can be realized as pure gauge modes in the massless vector perturbation, and Eq.~(\ref{VSLRES}) only becomes physical when $\mu \neq 0$. In Fig.~\ref{fig:effectivepotential}, one may observe that the effective potentials for the even-parity scalar modes converge to the Regge-Wheeler scalar potential in the massless limit, and the maximum of the effective potentials increases as the Proca mass parameter increases. The rate of increase is stronger than in the case of massive scalar perturbations, which can be confirmed by the coefficient of the $\mu^2$ term in Eq.~(\ref{VSLRESep}), while the coefficient for the massive scalar perturbation is one. Note that we also plot the Regge-Wheeler scalar potential and the effective potential for the massive scalar perturbation ($\mu=0.5$) in Fig.~\ref{fig:effectivepotential}, using the “pink-dotted” line and the “gray-dotted” line, respectively. Next, by comparing the same-colored “dashed” and “solid” lines in Fig.~\ref{fig:effectivepotential}, the maximum of the even-parity scalar potential is larger than that of the odd-parity one when all parameters are fixed. This observation can be further extended: the maxima of the even-parity scalar potentials are larger than those of the odd-parity ones when $l$ and $\mu$ are fixed and $\omega$ is varied, as shown in Fig.~\ref{fig:effectivepotentialfixlmu}, where the maximum of the effective potential for the even-parity scalar modes (dotted line) increases with increasing $\omega$. The asymptotic behavior of $V_{eff}^{(even-scalar)}(l=2;~\mu=0.4) \sim 0.4437$ can also be confirmed numerically for extremely large $\omega$, indicating that the full spectrum for this set of parameters remains barrier-like.

For the even-parity vector modes, the series expansion of Eq.~(\ref{VSLREV}) in the limit $\mu\rightarrow 0$ is
\begin{equation}
V_{eff}^{(even-vector)} \sim f \frac{l(l+1)}{r^2} \left[\frac{l^{2}(l+1)^{2}+\left(1-4 f\right)l(l+1)-r^2 \omega ^2}{l^{2}(l+1)^{2}+r^2 \omega ^2}+\frac{3 f l^{3}(l+1)^{3}}{\left(l^{2}(l+1)^{2}+r^2 \omega ^2\right)^2}\right]+ \mathcal{O}\left(\mu^{2}\right),\label{VSLREVep}
\end{equation}
 where the leading order is still $\omega$-dependent and does not exactly converge to the effective potential of the Regge-Wheeler vector equation. Nevertheless, the isospectrality of the quasi-normal frequencies in the Schwarzschild spacetime has been confirmed between the even-parity vector and odd-parity modes in the massless limit of the Proca radial equations (VSH separation \cite{DolanRosa2012}, FKKS separation \cite{PreDol2020}), and we will provide further discussion of this point based on our results. In Figs.~\ref{fig:effectivepotential} and \ref{fig:effectivepotentialfixlmu}, the maximum of the effective potentials for the even-parity vector modes tends to be the lowest compared to the odd-parity and even-parity scalar modes under the same parameter settings, except in the low-energy region of Fig.~\ref{fig:effectivepotentialfixlmu}. In addition, Fig.~\ref{fig:effectivepotential} also shows that, as the field becomes massive, the effective potentials for the even-parity vector and odd-parity modes deviate from the massless case (pink solid line), resulting in a lower and a higher peak value, respectively. Beyond the low-mass region, the effective potentials for all three modes tend to increase with increasing mass parameter. However, the relative order of the peak values—“even-parity scalar $>$ odd-parity $>$ even-parity vector”—remains consistent. Therefore, it is worth highlighting a characteristic feature of the even-parity vector modes. For fixed values of $l$ and $\omega$, the maximum of the effective potential initially decreases as the Proca mass increases in the small-mass region. Then, beyond a certain turning point of the Proca mass parameter, the maximum begins to increase as the mass continues to grow, until it reaches the barrier-like limit. Lastly, for the even-parity vector modes in Fig.~\ref{fig:effectivepotentialfixlmu}, the maximum of the effective potentials tends to decrease with increasing $\omega$. As the energy increases further, the effective potentials begin to deviate from the typical barrier-like structure and become dominated by potential wall features, which fall beyond the scope of the present study.
 
\section{Results by using the Rigorous bound method}\label{sec:bound}
\subsection{General formula for the rigorous bound method}
The rigorous bound method was developed to constrain general lower and upper bounds for the transmission and reflection probabilities of the time-independent Schrödinger equation \cite{PhysRevA.59.427}. The boundary conditions were chosen to resemble those of a one-dimensional quantum scattering problem, where, depending on the setup, a purely transmitted wave is imposed at one spatial infinity (either positive or negative), while a combination of incoming and reflected waves is present at the other. In 2008, Boonserm and Visser applied this method to derive bounds on the greybody factors of massless bosonic particles in the Schwarzschild black hole spacetime \cite{BooVi2008}. Further applications involving different types of particles in various black hole backgrounds have also been studied \cite{BoNgVi2014, Boonserm2017}.

The general form for bounding the greybody factor is given by
\begin{equation}
	T \geq \mathrm{sech}^{2}\int_{-\infty}^{\infty}\frac{\sqrt{\left[\partial_{r_{\star}}h(r_{\star})\right]^{2}+[\omega^2 -V_{eff}^{(modes)}-h^{2}(r_{\star})]^{2}}}{2h(r_{\star})} dr_{\star} , \label{bound}
\end{equation} 
where $h(r_{\star})$ is an arbitrary positive function related to the asymptotic behavior of the Schrödinger-like equation. In our case, which required to satisfy
\begin{equation}
h\left(\pm\infty\right)=\sqrt{\omega^{2}-V_{eff}^{(modes)}|_{r_{\star}\rightarrow\pm\infty}}.
\end{equation}
For studies of the perturbations of massive particles, it is natural to choose either $h\sim\sqrt{\omega^{2}-f\mu^2}$ or $h=\sqrt{\omega^{2}-V_{eff}^{(modes)}}$. The condition for an efficient (i.e., mathematically allowed) region is given by the positivity requirement, which corresponds to either $\omega\gtrsim\mu$ or $\omega\geq~\sqrt{V_{eff}^{(modes)}|_{max}}$, respectively. Since the latter always provides a stronger bound than the former \cite{BooVi2008}, we refer to the latter as the “strict bounds” and the former as the “less strict bounds” for convenience in the following discussions.

\subsection{Strict bounds}
We start with the strict bound by taking $h=\sqrt{\omega^{2}-V_{eff}^{(modes)}}$ which provides the strongest bound with the efficient region $\omega\geq~\sqrt{V_{eff}^{(modes)}|_{max}}$, and Eq.~(\ref{bound}) can be simplified as
\begin{equation}
	T \geq \mathrm{sech}^{2}\int_{-\infty}^{\infty}\frac{\sqrt{\left(\partial_{r_{\star}} h(r_{\star})\right)^{2}}}{2h(r_{\star})} dr_{\star}=\mathrm{sech}^{2}\int_{-\infty}^{\infty}\frac{|\partial_{r_{\star}} h(r_{\star})|}{2h(r_{\star})} dr_{\star}. \label{Sbound}
\end{equation}
The integral inside the bracket is a natural logarithm, and the sign depends on the first order derivative of the tortoise coordinate on $V_{eff}^{(modes)}$ since
\begin{equation}
\partial_{r_{\star}} h(r_{\star})=-\frac{\partial_{r_{\star}}V_{eff}^{(modes)}}{2h(r_{\star})}.
\end{equation} 
For the odd-parity mode, the radial positions of the local extrema of the effective potential are given by Eq.~(\ref{rmami}), and $r_{min} > r_{max}$ is always true for the barrier-like effective potentials with non-vanishing $l$. As such, the integral in Eq.~(\ref{Sbound}) shall be separated into three regions due to the absolute value in Eq.~(\ref{Sbound}), and the analytical solution for the strict bound of the odd-parity modes can be obtained.
\begin{equation}\label{AOB}
T\geq\frac{4\omega\sqrt{\omega^{2}-\mu^{2}}(\omega^{2}-V_{eff}^{(odd)}|_{max})(\omega^{2}-V_{eff}^{(odd)}|_{min})} {\left[(\omega^{2}-V_{eff}^{(odd)}|_{max})\sqrt{\omega^{2}-\mu^{2}}+\omega (\omega^{2}-V_{eff}^{(odd)}|_{min})\right]^{2}},
\end{equation}
where $V_{eff}^{(odd)}|_{max/min}\equiv V_{eff}^{(odd)}(r_{max/min})$. Note that the same mathematical expression as in Eq.~(\ref{AOB}) can be obtained for the strict bound of massive scalar particles in spherically symmetric black hole backgrounds \cite{polkong2025}, where the effective potential of massive scalar perturbation was given by \cite{KONOPLYA2005377}. For the even-parity modes, the effective potentials are $\omega$-dependent. Therefore, the analytical expressions of $r_{max/min}$ and $V_{eff}^{(even)}|_{max/min}$ cannot be obtained without series expansions. Nevertheless, numerical integration of Eq.~(\ref{Sbound}) is possible by fixing $l$ and $\mu$, and the resulting greybody factor will be a function of $\omega$.

\begin{figure}
    \centering
    \includegraphics[width=0.5\linewidth]{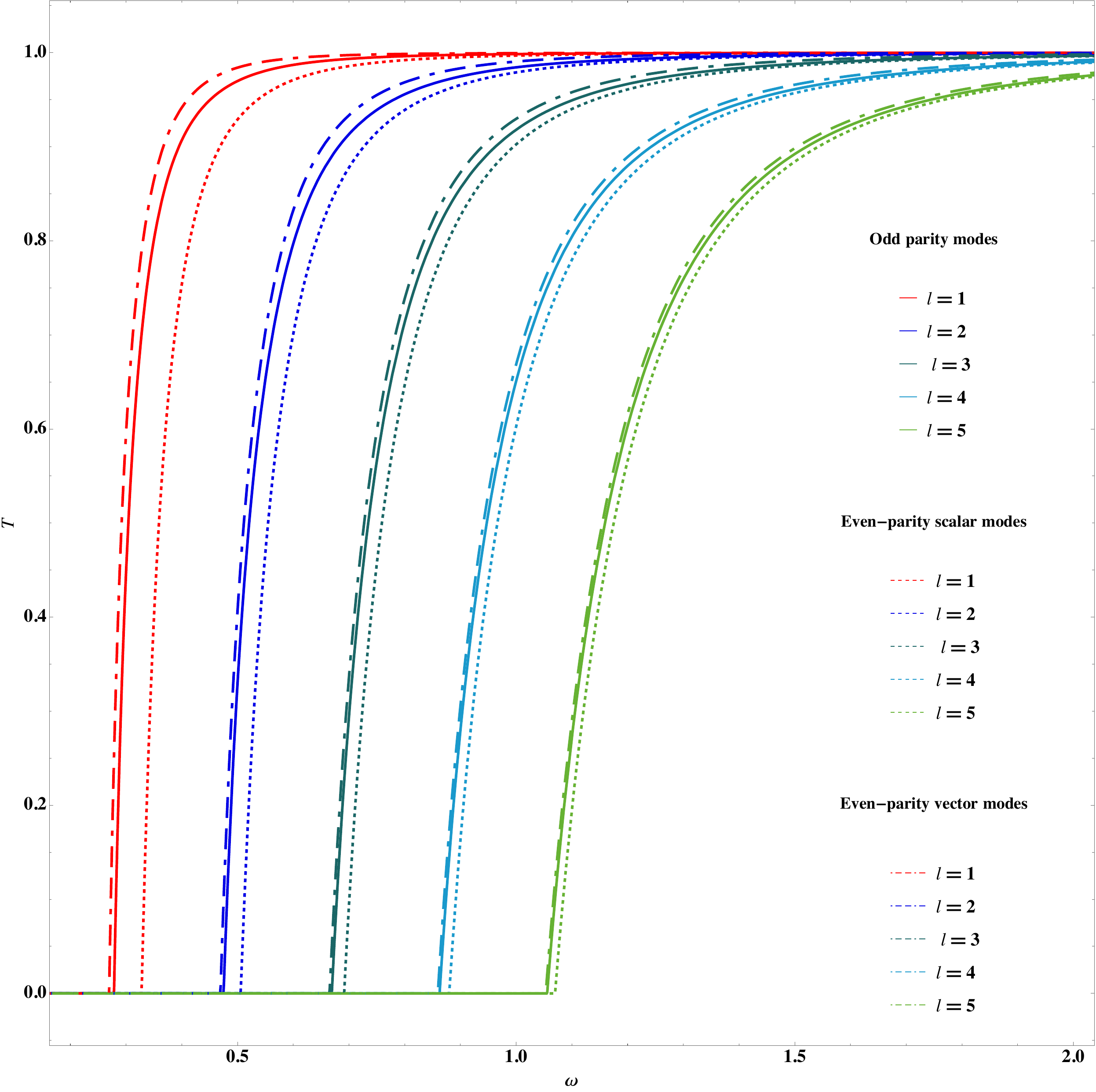}
    \caption{Comparison of the strict bounds for the three modes with fixed parameter $\mu=0.1$}
    \label{fig:sbl}
\end{figure}

In Fig.~\ref{fig:sbl}, we fix $\mu = 0.1$ and compare the results for the three modes with varying $l$, indicated by different colors. The solid line represents the odd-parity mode, while the dash-dotted and dotted lines represent the even-parity vector and even-parity scalar modes, respectively. Note that the plot style will be consistently applied to the other graphs in the article unless otherwise specified. We find that the order of the greybody factors using the strict bound follows the order of the effective potentials discussed in the previous subsection, where the maxima satisfy: even-parity scalar $>$ odd-parity $>$ even-parity vector. Accordingly, the greybody factors, from highest to lowest with fixed $l$ and a specified cutoff for the same $\omega$, follow the order: even-parity vector $>$ odd-parity $>$ even-parity scalar. As $l$ increases, the greybody factor curves tend to shift toward higher $\omega$. This behavior arises from the fact that a larger $l$ leads to a higher peak in the effective potential, as can be checked from Eqs.~(\ref{VSLRO}) to (\ref{VSLREV}).
 
\begin{figure}
    \centering
    \includegraphics[width=0.5\linewidth]{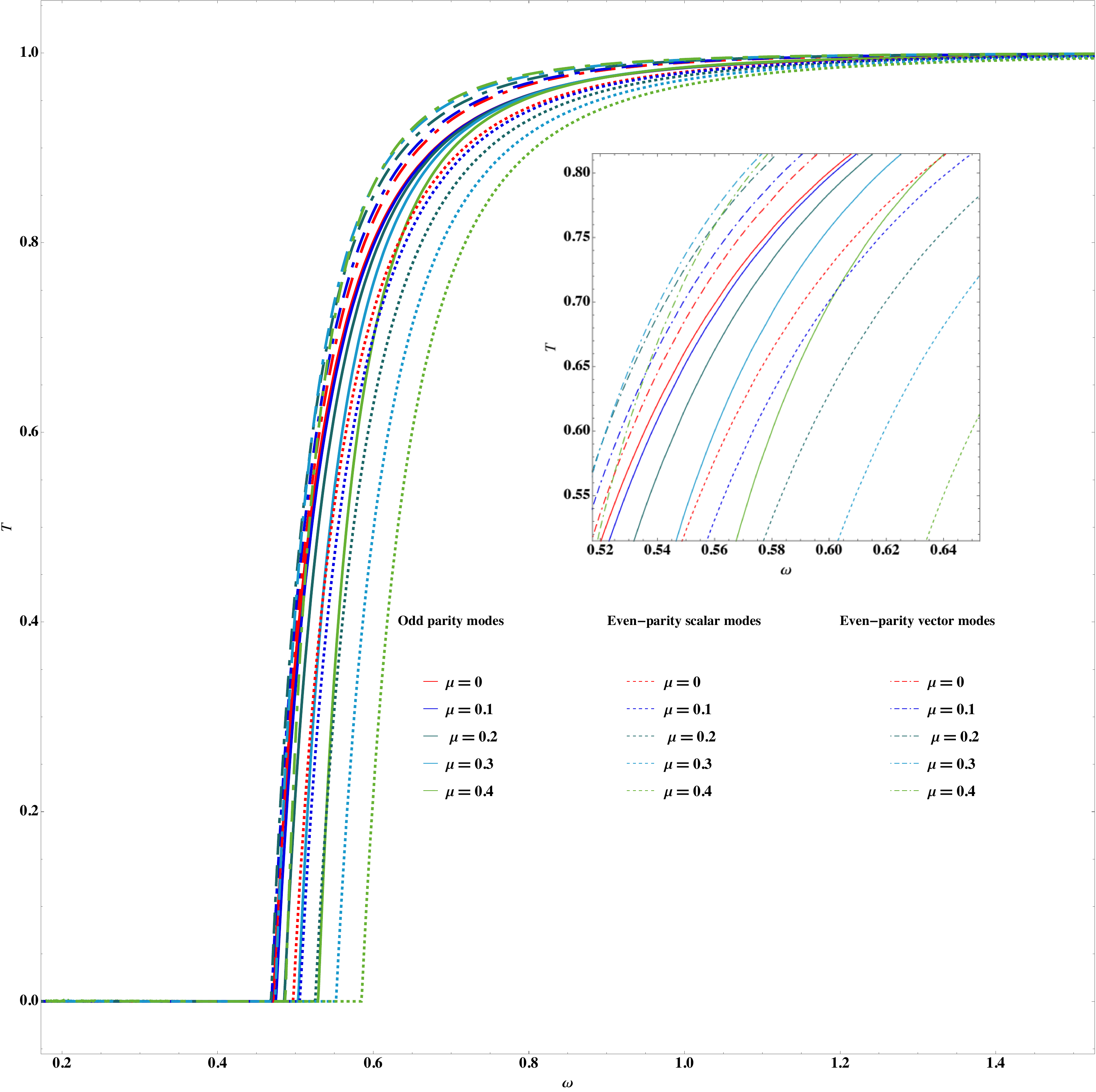}
    \caption{Comparison of the strict bounds for the three modes with fixed parameter $l=2$}
    \label{fig:sbmu}
\end{figure}

In Fig.~\ref{fig:sbmu}, we fix $l=2$ and compare the results for the three modes with varying $\mu$, with zoomed-in results for $0.55\lesssim\omega\lesssim 0.66$ and $0.55\lesssim T\lesssim 0.8$ presented in the sub-figure. For a fixed $\mu$ and the same value of $\omega$, the ordering of the greybody factors among all modes is consistent with that shown in Fig.~\ref{fig:sbl}. For the odd-parity modes, the massless limit corresponds exactly to the Regge-Wheeler vector result. As $\mu$ increases, the greybody factor shifts to the right. This can be understood as follows: for a fixed $\omega$, increasing $\mu$ leads to a lower transmission probability; conversely, for a fixed transmission probability (e.g., $0.5$), a higher $\mu$ requires a higher energy to achieve. For the even-parity scalar modes, the massless limit is exactly the Regge-Wheeler scalar result. However, for the Proca field, the massless limit in this case corresponds to a pure gauge mode, which only becomes physical when $\mu \neq 0$, as discussed in the previous section. The behavior of the greybody factor shifts to higher energy as $\mu$ increases, similar to what we observe for the odd-parity modes.

In contrast, for the even-parity vector modes, isospectrality with the odd-parity modes in the massless limit is not observed under the strict bound. We believe this is because the result obtained using this method relies on a mathematical inequality that yields a lower bound, implying that the upper-left region corresponds to the allowed region. That is, if isospectrality holds, the result for the odd-parity mode should approach the allowed region of the massless even-parity vector one, which we will examine further using the next method in this work. Next, as the Proca mass parameter $\mu$ increases, the greybody factor initially shifts toward the upper-left area, then turns and shifts to the right. This indicates that, for a certain $\omega$, there exists a turning value $\mu_{\textrm{tur.}}$ where the behavior of the greybody factor changes. As $\mu$ continues to increase, it approaches a critical value $\mu_{\textrm{cri.}}$, beyond which the transmission probability surpasses the massless case and continues moving rightward, eventually approaching the barrier-like effective potential limit. That is, for fixed $l$ and in the range $0<\mu<\mu_{\textrm{cri.}}$, the greybody factor can exceed that of the massless case (i.e., the photon). This turning behavior is consistent with the trend in the effective potentials shown in Fig.~\ref{fig:effectivepotential}, where the maximum of the effective potential first decreases—dropping below the massless case—then increases again, crossing over at $\mu_{\textrm{cri.}}$ and continuing toward the barrier-like limit.

Furthermore, both the turning and critical values depend on $\omega$. This can be seen from the sub-figure in Fig.~\ref{fig:sbmu}, where at $\omega = 0.52$, the greybody factor for $\mu = 0.4$ (light-green dash-dotted line) has already crossed the massless limit curve (red dash-dotted line) and is shifting rightward. In contrast, at $\omega = 0.56$, the greybody factor for $\mu = 0.4$ is just beginning to turn; the transmission probability is higher than that for the massless case and for $\mu = 0.1$ and $0.2$, but slightly lower than that for $\mu = 0.3$. That is, $\mu = 0.4$ can be considered the “critical value” around $0.52 < \omega < 0.53$, but corresponds only to the “turning value” at $0.56 \lesssim \omega$. The physical explanation for the “critical value” is supported by Fig.~\ref{fig:effectivepotentialfixlmu}. When $\omega = 0.5$ (dark green dash-dotted line), the maximum of the effective potential is higher than that of the massless case (pink solid line), resulting in a lower transmission probability. As the energy increases up to $\omega = 1$, the maximum of the effective potential falls below that of the massless case, leading to a higher greybody factor. Therefore, $\mu = 0.4$ serves as the “critical value” within the region $0.5 < \omega < 1$, which is consistent with our observation in Fig.~\ref{fig:sbmu}.

Lastly, in the case of strict bounds, a typical cutoff at a specific $\omega$ in the low transmission probability region depends on the effective range where $\omega \geq \sqrt{V_{eff}^{(modes)}|_{max}}$. In such a region, the bounds should be corrected using the less strict bound introduced in the next section, which yields a broader effective domain than the strict one.
\subsection{Less strict bounds}
\begin{figure}
\centering
	\begin{minipage}[t]{0.3\linewidth}
		\centering
        \includegraphics[width=4.5cm]{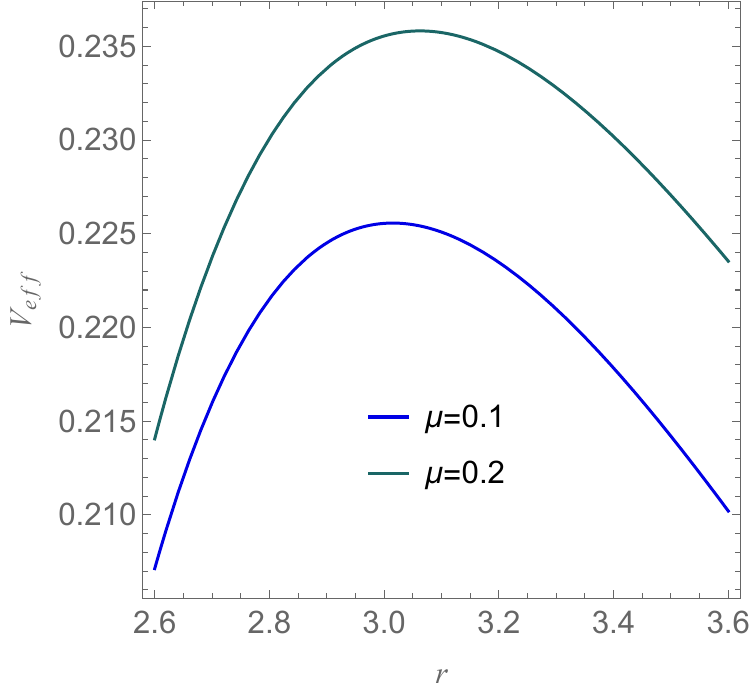}
	    \includegraphics[width=4.5cm]{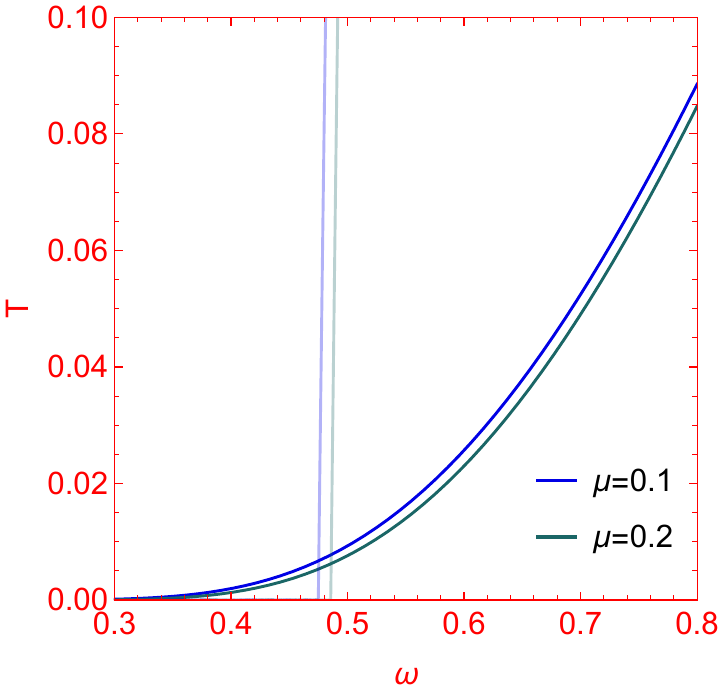}
        \subcaption{Odd-parity}
	\end{minipage}
	\begin{minipage}[t]{0.3\linewidth}
		\centering
		\includegraphics[width=4.5cm]{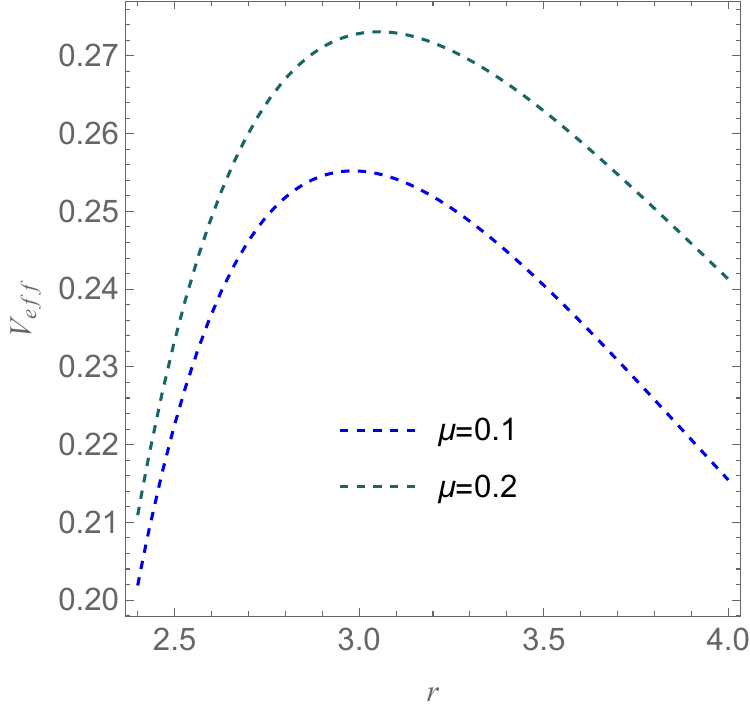}
        \includegraphics[width=4.5cm]{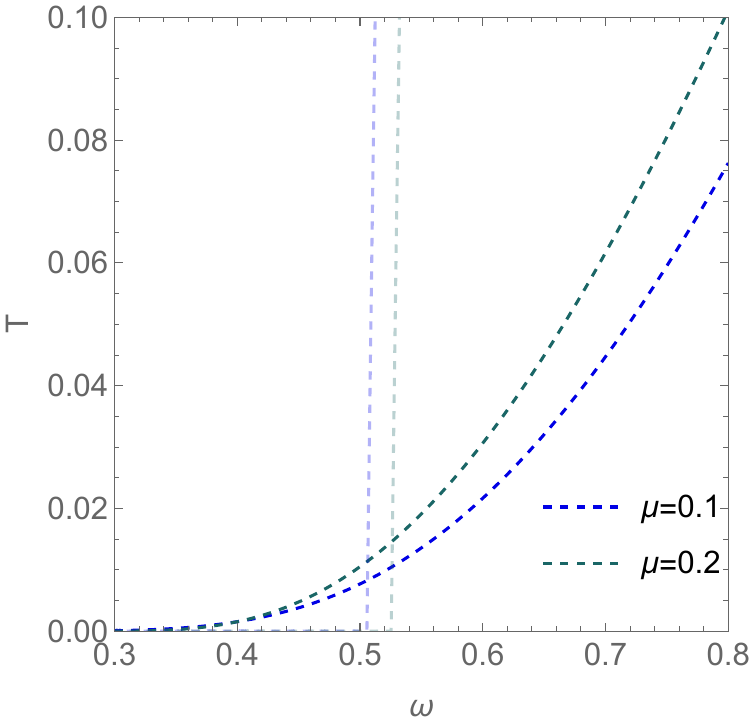}
        \includegraphics[width=4.5cm]{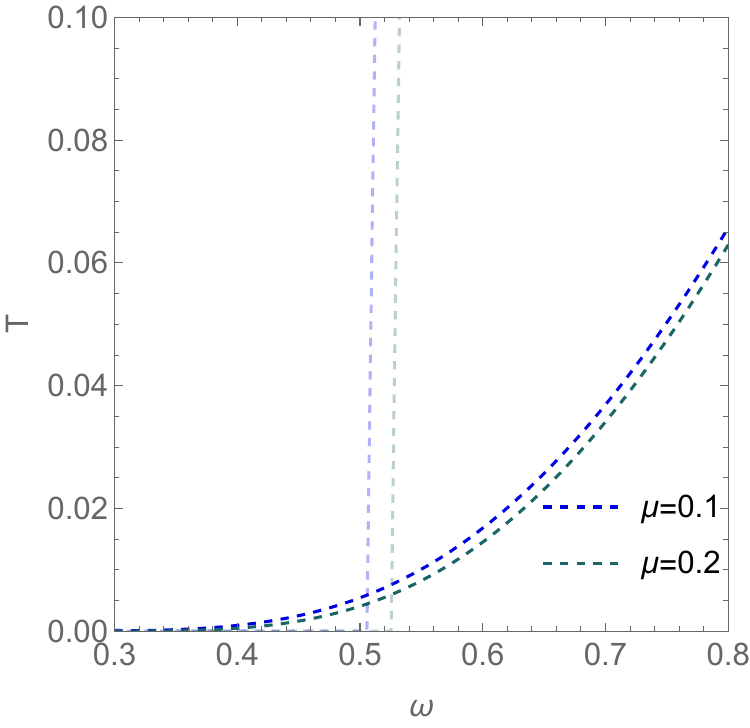}
        \includegraphics[width=4.5cm]{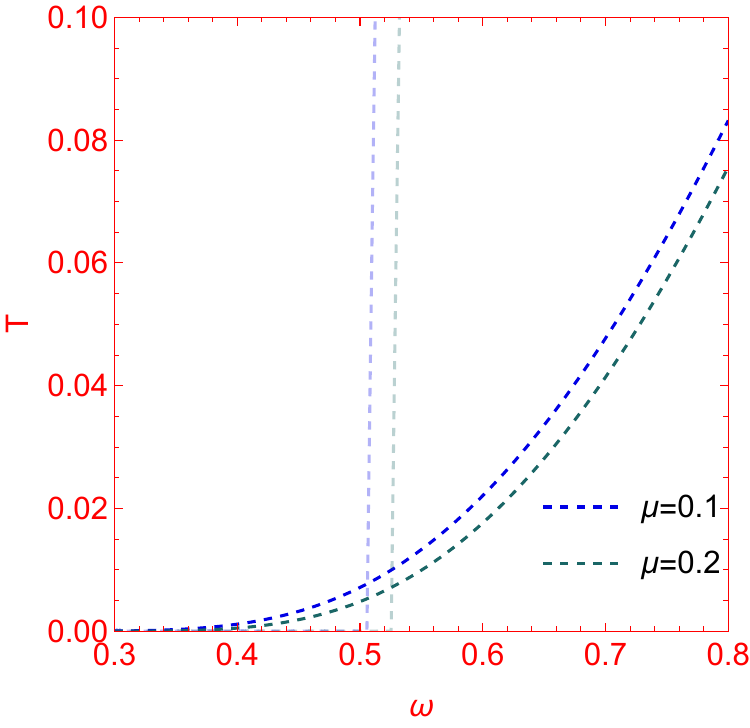}
        \subcaption{Even-parity scalar}
	\end{minipage}
	\begin{minipage}[t]{0.3\linewidth}
		\centering
		\includegraphics[width=4.5cm]{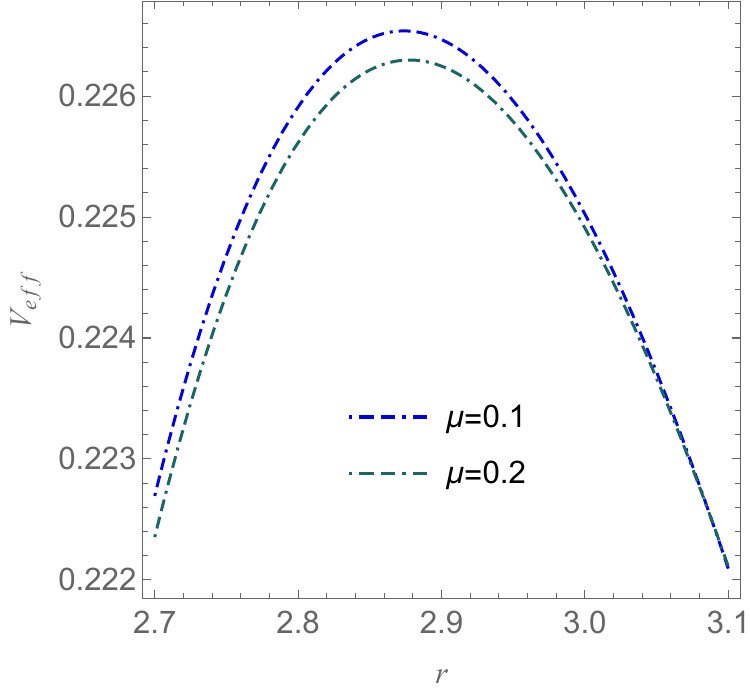}
        \includegraphics[width=4.5cm]{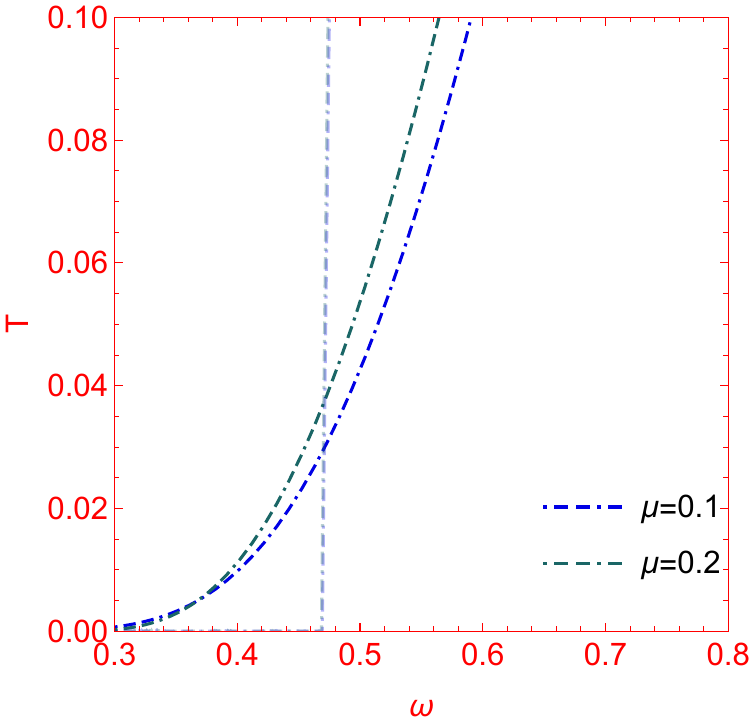}
        \includegraphics[width=4.5cm]{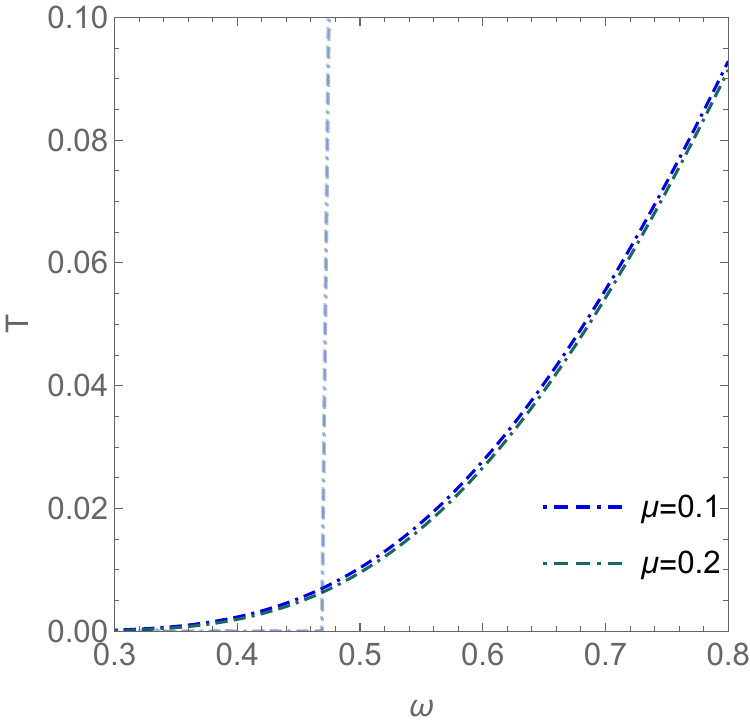}
        \includegraphics[width=4.5cm]{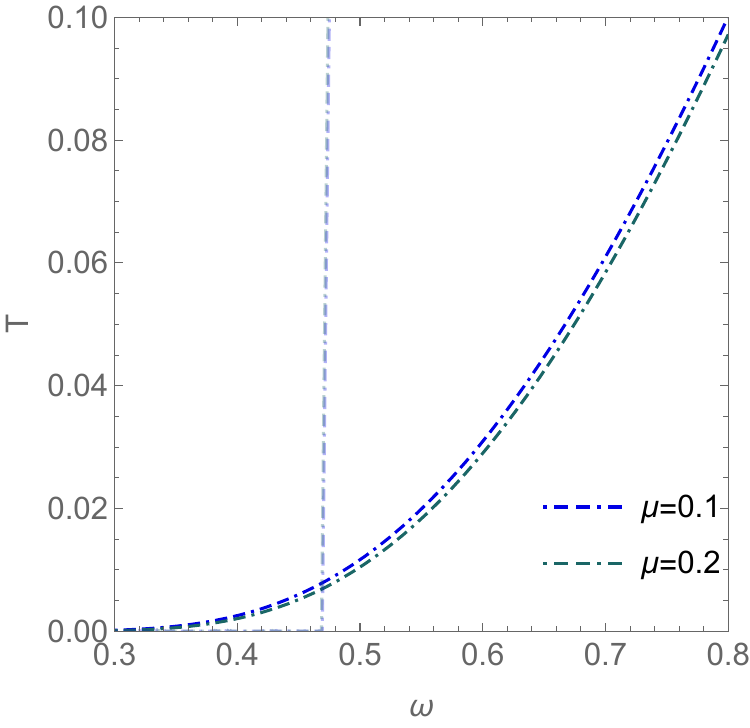}
        \subcaption{Even-parity vector}
	\end{minipage}
\caption{The low energy correction for three modes with $l=2$ by less strict bounds and the corresponding peak region of the effective potentials with $\omega=0.4$}\label{fig:lessstrictbounds}	
\end{figure}
In this case, taking $h = \sqrt{\omega^{2} - f \mu^2}$ together with the effective potential, Eq.~(\ref{VSLRO}), in Eq.~(\ref{bound}) provides a fundamental bound for the odd-parity case. However, this choice of $h$ fails to evaluate the less strict bound for the even-parity modes, as the numerical integration of Eq.~(\ref{bound}) tends to diverge at spatial infinity. Therefore, a modification of the $h$ function is necessary for the even-parity modes.

To rewrite the $h$ functions, we first consider the last term in Eqs.~(\ref{VSLRES}) and (\ref{VSLREV}), together with Eq.~(\ref{nunu}). From this, an explicit $\mu^{2}$-dependent term can be derived. The minimal requirement for modifying the $h$ function for the even-parity modes—so as to prevent the numerical integration of Eq.~(\ref{bound}) from diverging—is given by
\begin{equation}
h_{1}^{(\pm)} = \sqrt{\omega^{2} - f \mu^2 - 2 f \frac{\mu^{2}}{q_{r}^{(\pm)}}}.
\end{equation}
In the second row of Fig.~\ref{fig:lessstrictbounds}, we present the results for the less strict bounds by fixing $l=2$ and adopting $h$ and $h_{1}^{(\pm)}$ for the odd-parity and even-parity modes, respectively. Note that the lighter lines represent the results obtained from the strict bounds, which serve to efficiently constrain those from the less strict bounds. The colors of the lighter reference lines follow the same scheme as those used for the less strict bounds shown in the figure. The less strict bounds significantly improve efficiency in the low-energy region; that is, the allowed region for the lower bound of the greybody factors lies in the upper-left region of Fig.~\ref{fig:lessstrictbounds}. The boundary of each mode initially follows the less strict bound, but after intersecting with the strict bound, it is subsequently governed by the strict bound.

We notice that the behavior of the less strict bounds for the even-parity scalar modes in the second row of Fig.~\ref{fig:lessstrictbounds} shifts to the left as $\mu$ increases, which does not follow the general trend of the corresponding effective potentials as in the first row. As such, we propose an additional setting of the $h$ function for the even-parity modes, defined as
\begin{eqnarray}
h_{2}^{(\pm)} &=& \sqrt{\omega^{2} - f \mu^2 - 2 f \left(\frac{\mu^{2}-l(l+1)\nu_{\pm}^{2}}{q_{r}^{(\pm)}}\right)},\\
h_{3}^{(\pm)} &=& \sqrt{\omega^{2} - f \mu^2 - 2 f \left(\frac{\mu^{2}-l(l+1)\nu_{\pm}^{2}}{q_{r}^{(\pm)}}\right) -\left(\partial_{r_*}\frac{f}{rq_{r}^{(\pm)}}\right)-\left(\frac{f}{rq_{r}^{(\pm)}}\right)^2}.
\end{eqnarray}  
The less strict bounds obtained by adopting $h_{2}^{(\pm)}$ and $h_{3}^{(\pm)}$ for the even-parity modes are presented in the third and fourth rows of Fig.~\ref{fig:lessstrictbounds}, respectively. The previously observed left-shifting behavior of the even-parity scalar modes has been corrected to right-shifting, with the bounds from $h_{3}^{(\pm)}$ being slightly stronger than those from $h_{2}^{(\pm)}$. However, the even-parity vector modes exhibit a right-shifting behavior and do not follow the trend of the effective potentials shown in the first row. This discrepancy can be understood as a consequence of the dominant influence of the mathematical form of the $h$ functions in the low-energy region, where the energy lies below the maximum of the effective potential. In such cases, the less strict bounds are only weakly affected by the potential peak, making them more sensitive to the specific form of the chosen $h$ function.

Consequently, we identify the red-marked frames in Fig.~\ref{fig:lessstrictbounds} as representing the strongest and most reliable bounds. Their behavior as $\mu$ varies is consistent with that of the strict bounds, providing both numerical stability and physical consistency. We therefore consider these results to be the most accurate within this framework and adopt them as the final recommendation for the less strict bound formulation in this analysis.

\section{Results by using the WKB approximation}\label{sec:WKB}
\subsection{General formula for the WKB approximation}
The third-order Wentzel-Kramers-Brillouin (WKB) approximation was applied to obtain the quasinormal modes of black holes by Iyer and Will in 1987~\cite{IyerWill1987}, and the transmission and reflection coefficients using the fourth-order WKB method were further studied by Will and Guinn in 1988~\cite{WillGuinn1988}. Higher-order approximations, including the sixth order and beyond, combined with Padé approximants to improve convergence, were later developed by Konoplya and collaborators \cite{KonoWKB2003, Kono2019WKB}. These extensions significantly improved the accuracy of quasinormal mode calculations, but yielded only marginal improvements for scattering problems \cite{Kono2019WKB}. In addition, Ref.~\cite{CCDW2008} pointed out that even-order WKB approximations introduce uncertainties, such as the transmission probability dropping back to zero in the larger-energy region. Therefore, we adopt the third-order WKB approximation in our study.

The general formula of the greybody factor obtained by the third-order WKB approximation is given by
\begin{equation}
    \mathcal{T}=\frac{1}{1+e^{2S\left(\omega\right)}},\label{TWKB}
\end{equation}
where
\begin{eqnarray}
S(\omega)&=&\pi k^{1/2}\left[\frac{1}{2}z_0^2+\left(\frac{15}{64}b^2_3-\frac{3}{16}b_4\right)z_0^4+ \left(\frac{1155}{2048}b_3^4-\frac{315}{256}b_3^2b_4+\frac{35}{128}b_4^2+\frac{35}{64}b_3b_5-\frac{5}{32}b_6\right)z_0^6\right]\nonumber\\
&&+\pi k^{-1/2}\left[\frac{3}{16}b_4-\frac{7}{64}b_3^2 -\left(\frac{1365}{2048}b_3^4-\frac{525}{256}b_3^2b_4+\frac{85}{128}b_4^2+\frac{95}{64}b_3b_5-\frac{25}{32}b_6\right)z_0^2\right]\nonumber\\
&&+\mathcal{O}\left(k^{1/2}z_0^8,~k^{-1/2}z_0^4,~k^{-3/2}z_0^0\right).\label{SWKB}
\end{eqnarray}
The parameters corresponding to our notation are given by
\begin{eqnarray}
Q\left(r_{\star}\right)=\omega^{2}-V_{eff}^{(modes)}\ \ &;& \ \ z_{0}^{2}=\frac{-2Q\left(r_{max}\right)}{\partial_{r_{\star}}^{2}Q\left(r_{\star}\right)|_{r\rightarrow r_{max}}}; \nonumber\\
k=\frac{1}{2}\partial_{r_{\star}}^{2}Q\left(r_{\star}\right)|_{r\rightarrow r_{max}}\ \ &;& \ \ b_{n}=\frac{1}{n!k} \partial_{r_{\star}}^{n}Q\left(r_{\star}\right)|_{r\rightarrow r_{max}},\label{bWKB}
\end{eqnarray}
where $r_{max}$ is the radial position corresponding to the maximum peak of the effective potential. For odd-parity modes, $r_{max}$ takes the form given in Eq.~(\ref{rmami}) when $\mu \neq 0$, and $r_{max} = 3$ when $\mu = 0$. The greybody factor for the odd-parity modes can be computed straightforwardly by substituting a given set of $l$ and $\mu$ into Eq.~(\ref{VSLRO}), together with Eq.~(\ref{rmami}), and then proceeding step by step using Eqs.~(\ref{bWKB}), (\ref{SWKB}), and (\ref{TWKB}). For even-parity modes, since the effective potentials depend on $\omega$, the precise value of $r_{max}$ also depends on the energy of the Proca particles. A more detailed analysis of the WKB approximation for the greybody factor of even-parity modes will be presented in the following subsections.
\subsubsection{Even-parity scalar modes}
As mentioned in Eq.~(\ref{VSLRESep}), the leading-order term in the series expansion of the effective potential, Eq.~(\ref{VSLRES}), as $\mu \rightarrow 0$, is exactly the effective potential of the Regge–Wheeler scalar equation. The corresponding $r_{max}$ at leading order, denoted as $r_0$, can be solved as
\begin{equation}
r_0(l) = \frac{-3 + 3l(l+1) + \sqrt{l(l+1)\left(9l(l+1)+14\right) + 9}}{2l(l+1)},\label{rmax0}
\end{equation}
which is $\omega$ independent and be a specific radial position with a given $l$. Therefore, one may rewrite the effective potentials and $r_{\text{max}}$ as power series in $\mu$, which allows us to solve for $r_{\text{max}}$ order by order. This procedure then proceeds to the WKB formula, through which the greybody factors can be obtained as a series in $\mu$. The idea of applying such a systematic expansion to the WKB approximation was first introduced by Simone and Will in their study of massive scalar quasinormal modes in the Kerr black hole \cite{LESimone_1992}, and later extended to study the quasinormal modes for the linear perturbation of the massive Dirac field in \cite{Cho2003}. A similar approach to analyzing greybody factors using this supplemental expansion was also employed for massive Dirac perturbations in a massive gravity black hole \cite{BoChNgW2021}. Accordingly, Eq.~(\ref{VSLRES}) and the corresponding $r_{max}$ can be expressed as
\begin{eqnarray}
V(\omega, \mu, l, r) &=& V_0(r, l) + V_1(\omega, l, r)\mu + V_2(\omega, l, r)\mu^2 + \cdots + V_n(\omega, l, r)\mu^n,\label{VEx}\\
r_{max} &=& r_0(l) + r_1(\omega, l)\mu + r_2(\omega, l)\mu^2 + \cdots + r_n(\omega, l)\mu^n \equiv r_0 + \delta,\label{rmax}
\end{eqnarray}
where $V_n(\omega, l, r)$ and $r_n(\omega, l)$ denote the coefficients at order $\mu^n$. We further assume that the higher-order corrections to $r_{max}$ can be naturally regarded as a small quantity $\delta$, indicating that the radial position of the peak deviates only slightly from the massless scalar case when $\mu$ is small. The condition for $r_{max}$ is given by
\begin{equation}
    0=\partial_rV(\omega,\mu,l,r)|_{r\rightarrow r_{\max}}.\label{ESmaxcondi}
\end{equation}
With a further expansion of $V(\omega,\mu,l,r)$ around $r_{max}$ and by substituting it into Eq.~(\ref{ESmaxcondi}), we get
\begin{eqnarray}
0&=&\partial_{r}V\left(\omega,\mu,l,r_{0}\right)+\delta \partial_{r}^{2}V\left(\omega,\mu,l,r_0\right)+\frac{1}{2}\delta^2\partial_{r}^{3}V\left(\omega,\mu,l,r_0\right) +\frac{1}{6}\delta^3\partial_{r}^{4}V\left(\omega,\mu,l,r_0\right)\nonumber \\
&&+\frac{1}{24}\delta^4\partial_{r}^{5}V\left(\omega,\mu,l,r_0\right)+\frac{1}{120}\delta^5\partial_{r}^{6}V\left(\omega,\mu,l,r_0\right) +\frac{1}{720}\delta^6\partial_{r}^{7}V\left(\omega,\mu,l,r_0\right),\label{Vr0}
\end{eqnarray}
where $\partial_{r}^{n}V\left(\omega,\mu,l,r_0\right)=\partial_{r}^{n}V\left(\omega,\mu,l,r\right)|_{r\rightarrow r_{0}}$. Substituting Eqs.~(\ref{rmax0}), (\ref{VEx}) and (\ref{rmax}) in Eq.~(\ref{Vr0}), giving a integer for $l$, collecting and solving the coefficients of $\mu$ as
\begin{eqnarray}
0 &= \mu &\left[\partial_{r}V_1(\omega, r_0) + r_1 \partial_{r}^{2}V_0(\omega, r_0)\right], \nonumber \\
0 &= \mu^2 &\left[\partial_{r}V_2(\omega, r_0) + r_2 \partial_{r}^{2}V_0(\omega, r_0) + r_1 \partial_{r}^{2}V_1(\omega, r_0) + \frac{1}{2} r_1^2 \partial_{r}^{3}V_0(\omega, r_0)\right], \nonumber \\
0 &= \mu^3 &\left.\bigg[\partial_{r}V_3(\omega, r_0) + r_3 \partial_{r}^{2}V_0(\omega, r_0) + r_2 \partial_{r}^{2}V_1(\omega, r_0) + r_1 \partial_{r}^{2}V_2(\omega, r_0) \right.\nonumber\\
&&\left.+ r_1 r_2 \partial_{r}^{3}V_0(\omega, r_0) + \frac{1}{2} r_1^2 \partial_{r}^{3}V_1(\omega, r_0) + \frac{1}{6} r_1^3 \partial_{r}^{4}V_0(\omega, r_0)\right].\\\label{ri}
&&\vdots\nonumber
\end{eqnarray}
For the coefficient of $\mu$ in Eq.~(\ref{ri}), where $V_1(\omega, r)=0$ is obviously when we compare Eq.~(\ref{VEx}) with Eq.~(\ref{VSLRESep}), and therefore $r_{1}=0$. By solving for $r_n$ order by order with respect to the coefficients of $\mu^n$, we can approximate $r_{\max}$ in the form given by Eq.~(\ref{rmax}). As examples, the analytical expressions for $l=1$ and $l=2$ are presented in Eqs.~(\ref{esrmaxl1}) and (\ref{esrmaxl2}), respectively.
\begin{eqnarray}
r_{max}(l=1)&=&2.886+8.44013 \mu ^2+\left(57.3746\, -\frac{17.5217}{\omega ^2}\right)\mu ^4 + \left(493.228\, -\frac{349.415}{\omega ^2}+\frac{70.0868}{\omega ^4}\right)\mu ^6,\label{esrmaxl1}\\
r_{max}(l=2)&=&2.95171+3.77813 \mu ^2+\left(10.3137\, -\frac{7.92111}{\omega ^2}\right)\mu ^4+\left(35.275\, -\frac{90.4458}{\omega ^2}+\frac{31.6844}{\omega ^4}\right)\mu ^6  .\label{esrmaxl2}
\end{eqnarray}
\begin{table}[t!]
\centering
\caption{The comparison for the numerical and approximated $r_{max}$ of even-parity scalar modes with different values of $\omega$, $\mu$ and $l=1,~2$.}\label{ESrpeak}
\begin{tabular}{cccccccc}
\hline
 l=1& & & &l=2 & & & \\
\hline
$\mu$ & $\omega$ & Numerical & Eq.~(\ref{esrmaxl1}) & $\mu$ & $\omega$ & Numerical & Eq.~(\ref{esrmaxl2}) \\
\hline
0.1 & 0.3 & 2.95250 & 2.98418  &0.1 & 0.3 & 2.98052 & 2.99561\\
0.1 & 0.6 & 2.96672 & 2.96740  &0.1 & 0.6 & 2.98646 & 2.98683\\
0.1 & 0.9 & 2.97157 & 2.97161  &0.1 & 0.9 & 2.98846 & 2.98849\\
0.1 & 1.2 & 2.97364 & 2.97363  &0.1 & 1.2 & 2.98931 & 2.98932\\
0.2 & 0.3 & 3.10522 & 5.96460  &0.2 & 0.3 & 3.04032 & 4.40453\\
0.2 & 0.6 & 3.19827 & 3.28976  &0.2 & 0.6 & 3.07241 & 3.12240\\
0.2 & 0.9 & 3.25589 & 3.26297  &0.2 & 0.9 & 3.09053 & 3.09562\\
0.2 & 1.2 & 3.29100 & 3.28819  &0.2 & 1.2 & 3.10092 & 3.10180\\
\hline
\end{tabular}
\end{table}

Note that the above formulas for $r_{\max}$ break down as $\omega \rightarrow 0$, and they also fail to accurately describe the behavior at small $\omega$ values, where higher-order terms in $\mu$ contribute more significantly than the lower-order terms. Therefore, the approximations in Eqs.~(\ref{esrmaxl1}) and (\ref{esrmaxl2}) are efficient primarily for small $\mu$ and relatively large $\omega$, as illustrated in Table~\ref{ESrpeak}. Since the maximum of the effective potential increases with both $\mu$ and $\omega$, as shown in Figs.~\ref{fig:effectivepotential} and~\ref{fig:effectivepotentialfixlmu}, the region of interest for the greybody factor overlaps with this efficient regime, which will be discussed in the next subsection.

By expressing $r_{max}$ as a series expansion and expanding the effective potential in Eq.~(\ref{VSLRES}) according to the structure of Eq.~(\ref{VEx}), we can apply Eqs.~(\ref{bWKB}), (\ref{SWKB}), and (\ref{TWKB}) to compute the greybody factors of even-parity scalar modes using the WKB approximation. It is important to note that, throughout this process, the series in $\mu$ must be reorganized to maintain consistency of the overall expansion up to order $\mu^6$, since $r_{\max}$ is expanded only up to this order.

\subsubsection{Even-parity vector modes}\label{sec:EPV}
For the even-parity vector modes, the mathematical structure of the effective potential in Eq.~(\ref{VSLREV}) is similar to that of the even-parity scalar modes in Eq.~(\ref{VSLRES}); the only difference lies in the ``$+$'' sign in Eq.~(\ref{nu}). This difference leads to a distinct convergence behavior of the parameter $\nu$ in the massless limit, and the leading-order term in the series expansion for the even-parity vector modes remains $\omega$-dependent, as shown in Eq.~(\ref{VSLREVep}). The leading-order value of $r_{max}$ for the even-parity vector modes cannot be determined independently of $\omega$. Therefore, it is more efficient to treat and evaluate $r_{max}$ separately for each $\omega$, as demonstrated in Table~\ref{rpeak}, where we provide examples of $r_{max}$ for the effective potential in Eq.~(\ref{VSLREV}) with $l = 2$ and Proca mass values ranging from 0.1 to 0.4.
\begin{table}[t!]
\centering
\caption{The $r_{max}$ of even-parity vector modes with different values of $\omega$, $\mu$ and $l=2$.}\label{rpeak}
\begin{tabular}{cccccc}
\hline
$\omega$ &~$\mu=0.1$~&~$\mu=0.2$~&~$\mu=0.3$~&~$\mu=0.4$~&~$\mu=0.5$~ \\
\hline
0.1 & 2.94550 & 2.96301 & 3.01012 & 3.10282 & 3.27622\\
0.2 & 2.92632 & 2.93681 & 2.97874 & 3.06826 & 3.24192\\
0.3 & 2.90222 & 2.90825 & 2.94663 & 3.03398 & 3.20859\\
0.4 & 2.87383 & 2.87775 & 2.91402 & 3.00006 & 3.17622\\
\hline
\end{tabular}
\end{table}

By combining each $r_{max}$ with Eqs.~(\ref{bWKB}), (\ref{SWKB}), and (\ref{TWKB}), the greybody factor can be obtained as a data point for each fixed set of parameters, and each point only represent the corresponding greybody factor for a specific particle energy $\omega$.  To obtain a full spectrum of the greybody factors for this mode, we fix a given set of $l$ and $\mu$, and begin from $\omega^2 = 0.005$, computing the corresponding $r_{max}$ and greybody factor. We then increment $\omega^2$ by steps of $0.005$, repeating the process continuously until the greybody factor approaches $1$. Finally, we connect all the data points to represent the result for the even-parity vector modes using the WKB approximation.
\subsection{WKB results}
\subsubsection{General comparison}
\begin{figure}
    \centering
    \includegraphics[width=0.5\linewidth]{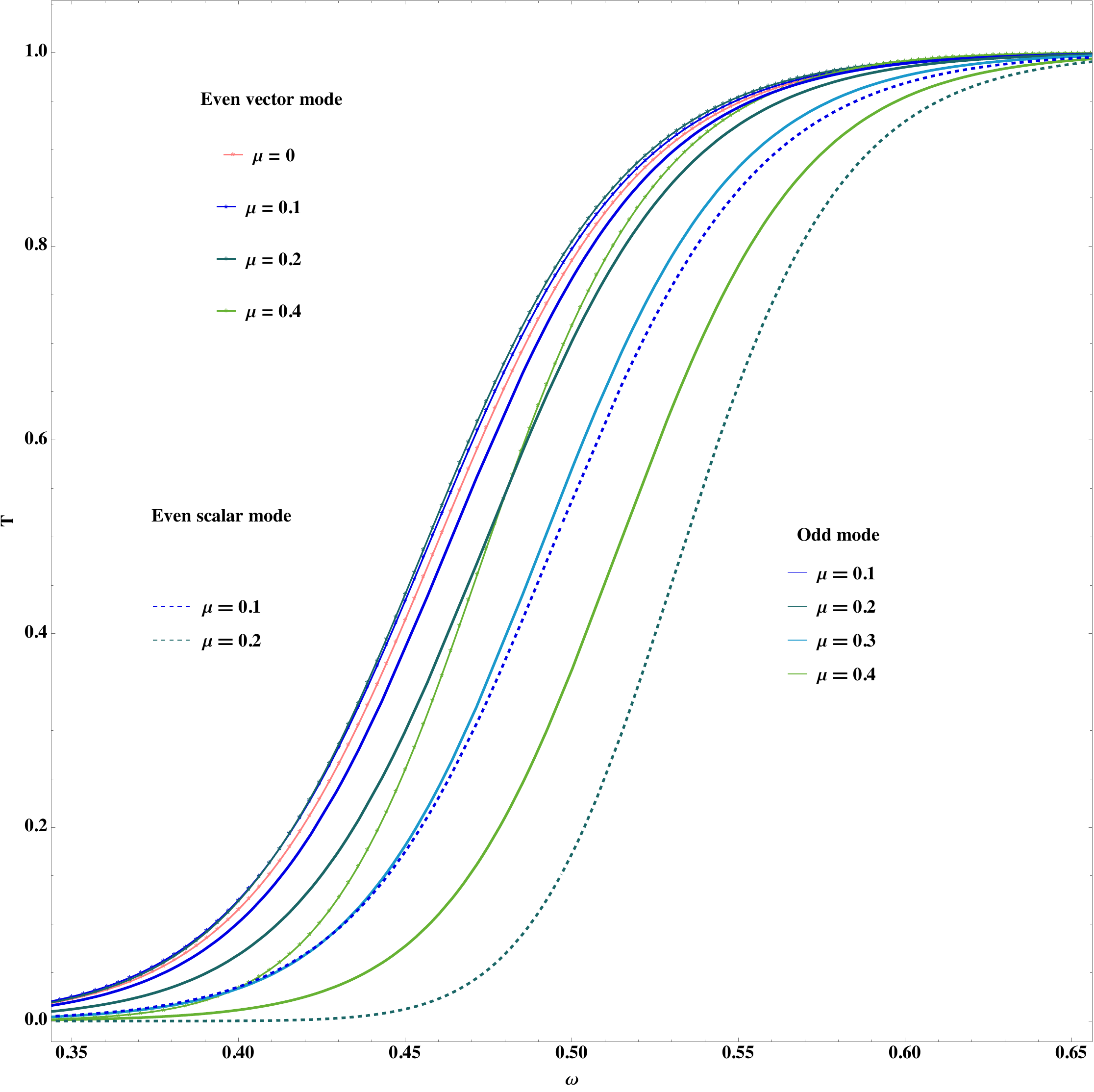}
    \caption{ The comparison for the greybody factor of the odd-parity, even-parity vector and scalar modes with $l = 2$ and varying $\mu$.}
    \label{fig:WKB3modes}
\end{figure}
\begin{figure}
\centering
	\begin{minipage}[t]{0.3\linewidth}
		\centering
        \includegraphics[width=4.5cm]{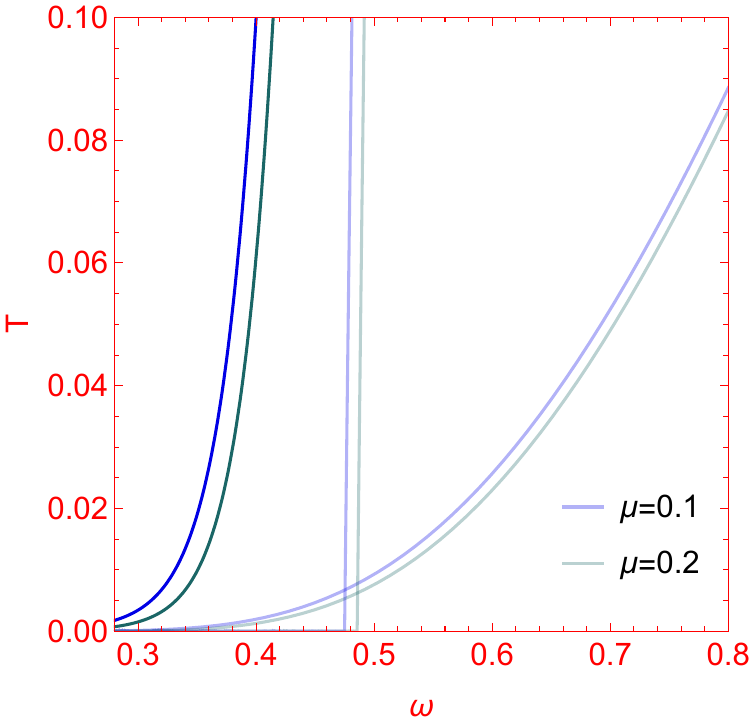}
        \subcaption{Odd-parity}
	\end{minipage}
	\begin{minipage}[t]{0.3\linewidth}
		\centering
		\includegraphics[width=4.5cm]{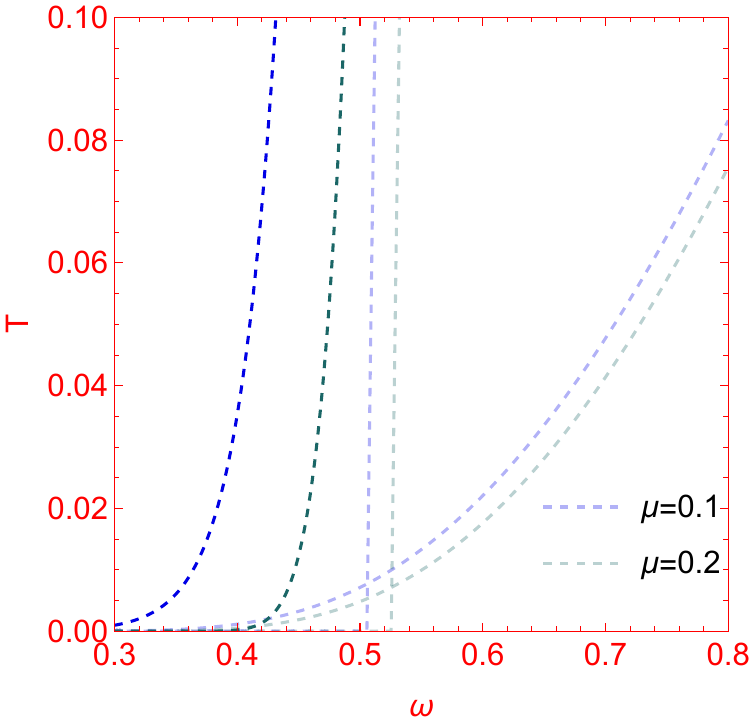}
        \subcaption{Even-parity scalar}
	\end{minipage}
	\begin{minipage}[t]{0.3\linewidth}
		\centering
		\includegraphics[width=4.5cm]{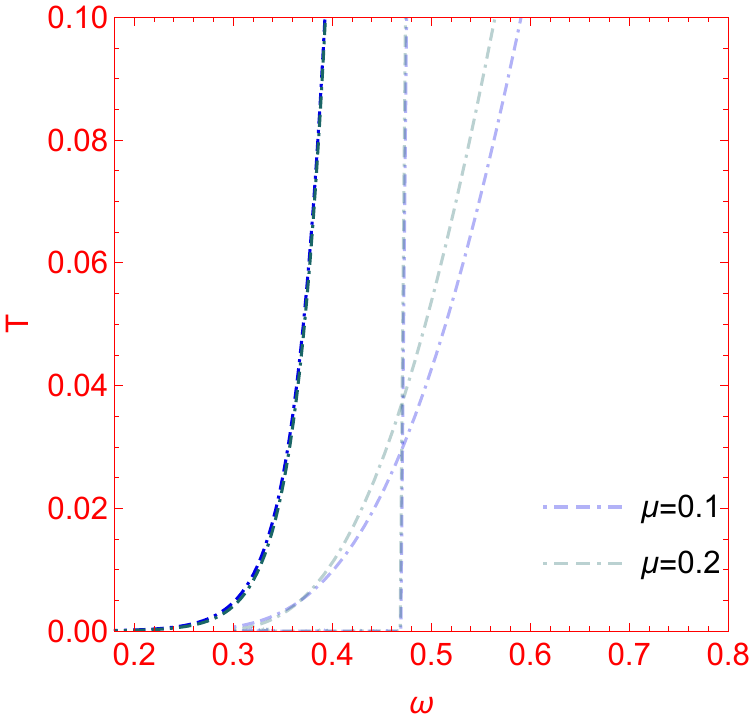}
        \subcaption{Even-parity vector}
	\end{minipage}
\caption{The consistency of the WKB results with the Rigorous bound method, where $l=2$ for all modes.}\label{fig:WKBRG}	
\end{figure}
The general comparison of the greybody factors for odd-parity, even-parity vector, and scalar modes obtained using the WKB approximation is presented in Fig.~{\ref{fig:WKB3modes}}, with $l = 2$ and varying $\mu$. As usual, the solid line and the dotted line represent the odd-parity mode and the even-parity scalar mode, respectively. The solid line with dots replaces the dash-dotted line to represent the results for the even-parity vector modes, where the dots indicate the actual data points, as discussed in the previous subsection.

The structure of the greybody factor for each mode is consistent with the results obtained from the rigorous bounds, although the overall profile shifts toward a lower-energy region, which is identified as the allowed region by the rigorous bounds. Specifically, for fixed $l$, $\omega$ and $\mu$, the greybody factors follow the order: even-parity vector $>$ odd-parity $>$ even-parity scalar. Alternatively, if we take the observation of fixing $l$, $\mu$, and the transmission probability (e.g., $T = 0.5$), the ordering from left to right remains: even-parity vector, odd-parity, and even-parity scalar. As $\mu$ increases, the greybody factor curves for the odd-parity and even-parity scalar modes shift to the right, which is also consistent with the behavior of the effective potentials shown in Figs.~\ref{fig:effectivepotential} and~\ref{fig:effectivepotentialfixlmu}, as well as the strict bounds illustrated in Fig.~\ref{fig:sbmu}.

The red solid line with dots, generated from the even-parity vector mode in the massless limit, also represents the result for the massless vector perturbation. From massless to massive vector perturbations, a clear splitting in the greybody factors between odd- and even-parity vector modes can be observed, with the former shifting toward higher-energy regions (rightward) and the latter toward lower ones (leftward), at fixed transmission probability $T$. With fixed $T$ and increasing $\mu$, the existence of a turning point (where the shift direction reverses) and a critical point (where the curve matches the massless case) for the even-parity vector modes naturally appears, as shown in the figure. It is worth noting that the result for the even-parity vector mode with $\mu = 0.3$ is omitted in Fig.~\ref{fig:WKB3modes} for the sake of clarity. One can expect that the $\mu = 0.3$ curve lies between the $\mu = 0.2$ and $\mu = 0.4$ curves and intersects several of the other curves.

The consistency with the rigorous bound can be checked by comparing Fig.~\ref{fig:WKB3modes} with Fig.~\ref{fig:sbmu}. As an example for the even-parity scalar mode with $l=2$, $\mu=0.2$, and $\omega\sim 0.6$, the resulting greybody factor from WKB is $T\sim 0.9$, and the lower bound is located at $T\gtrsim 0.64$. Therefore, the WKB result of this mode satisfies the constraint imposed by the lower bound for this particular set of $\{l,\mu,\omega\}$. This comparison can be extended to all the results. For the region of lower transmission probability, the consistency becomes clearer through Fig.~\ref{fig:WKBRG}, where the WKB results are plotted in pure colors and the bounds are the red-framed selections from Fig.~\ref{fig:lessstrictbounds}, which appear in lighter colors. Therefore, we may summarize that the WKB results are consistent with the lower bounds in our study.

\subsubsection{Isospectrality in the massless limit}\label{sec:Iso}
\begin{figure}
    \begin{subfigure}{0.45\textwidth}
    \includegraphics[width=\textwidth]{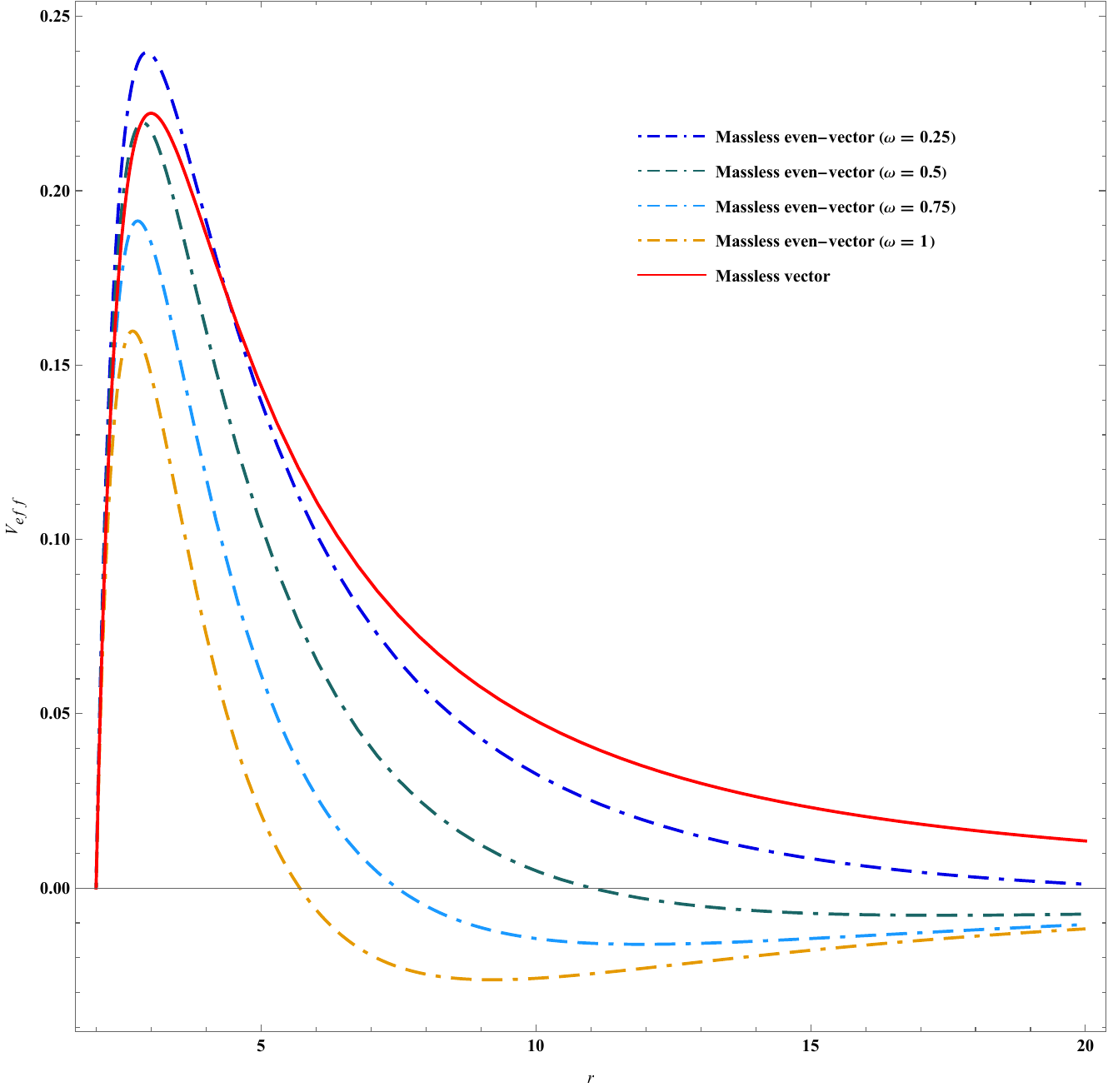}
    \caption{Effective potential.}\label{ISOV}
    \end{subfigure}
    \begin{subfigure}{0.45\textwidth}
    \includegraphics[width=\linewidth]{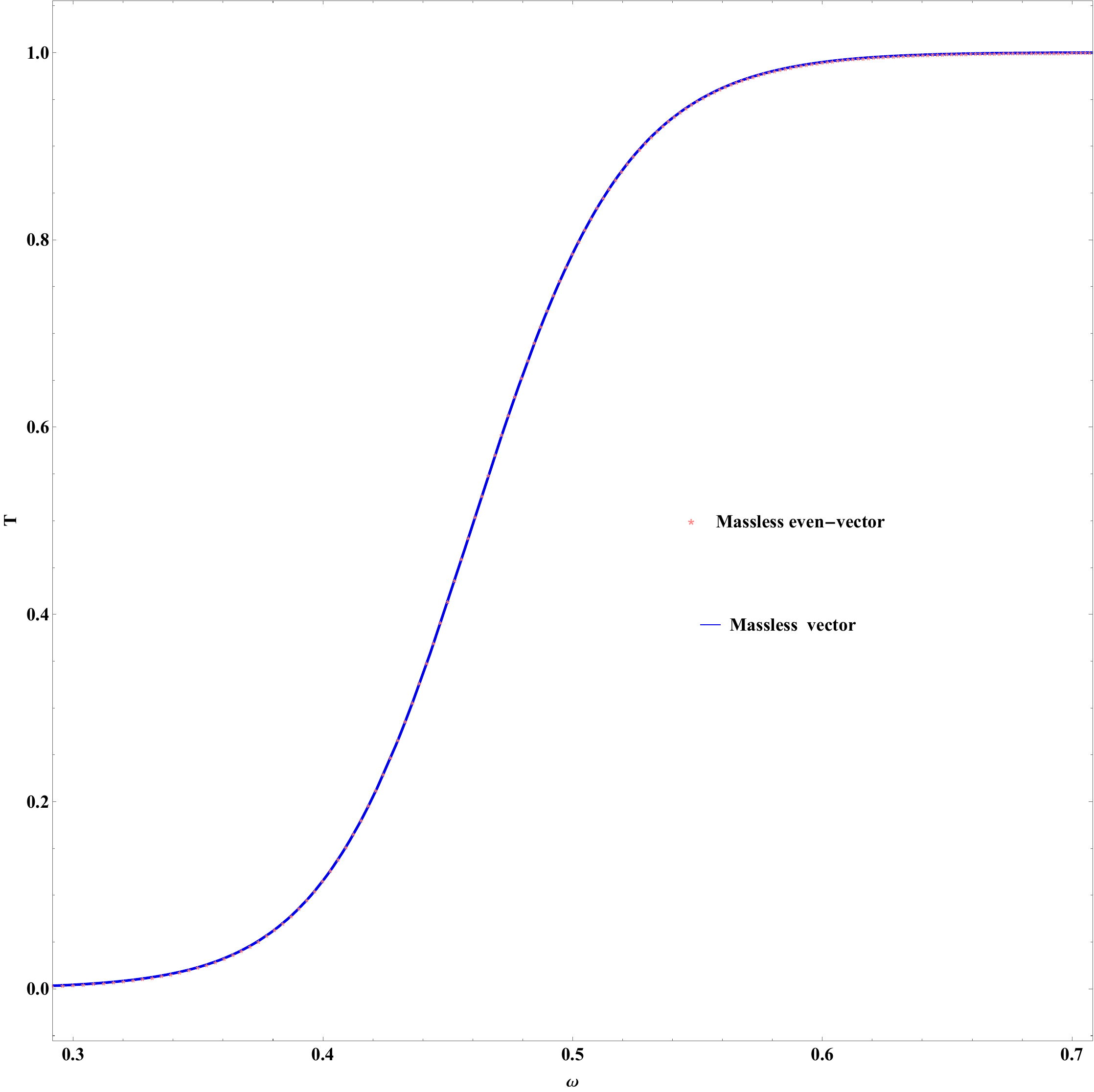}
    \caption{Greybody factor}\label{ISOT}
    \end{subfigure}
\caption{Isospectrality between the even-parity vector and odd-parity modes in the massless limit with $l = 2$.}
\label{fig:isoevod}
\end{figure}
\begin{table}[t!]
\centering
\caption{The greybody factors for the massless limit of the odd-parity and even-parity vector modes with $l=2$.}\label{Isisoectral}
\begin{tabular}{ccccccccccc}
\hline
    &$\omega=0.1$ & $\omega=0.2$ & $\omega=0.3$ & $\omega=0.4$ & $\omega=0.5$ & $\omega=0.6$ & $\omega=0.7$ & $\omega=0.8$ & $\omega=0.9$ & $\omega=1.0$ \\
\hline
Eq.~(\ref{Toddmassless}) & 0.00002 & 0.00021 & 0.00438 & 0.11558 & 0.78529 & 0.98972 & 0.99982 & 1.00000 & 1.00000 & 1.00000\\
even-parity vector     & 0.00002 & 0.00019 & 0.00422 & 0.11527 & 0.78517 & 0.98953 & 0.99980 & 0.99999 & 0.99999 & 1.00000\\
\hline
\end{tabular}
\end{table}

An important result for massless vector perturbations in the Schwarzschild black hole is that the even-parity and odd-parity modes share the same radial equation, which takes the Regge–Wheeler form. Upon introducing the Proca mass, the vector-type polarization splits into odd-parity and even-parity vector modes. The degeneracy of physical results in the massless limit of Proca perturbations in the Schwarzschild black hole has been confirmed by numerically solving the coupled radial equations for quasinormal modes in \cite{DolanRosa2012} and for the transmission probability in \cite{HerSamWan2012}. Further confirmation was provided in \cite{PreDol2020}, which reproduced the isospectrality of quasinormal modes by using the radial equation derived from the FKKS ansatz in the static and massless limit. 

To check the isospectrality, the greybody factor for the odd-parity modes in the massless limit with $l=2$ is given by
\begin{equation}
T_{\text{odd}}|_{\mu\rightarrow 0}=\left[1+e^{\frac{\pi \left(-6274503 \omega ^6+6597450 \omega ^4-3696750 \omega ^2+546988\right)}{62208 \sqrt{6}}}\right]^{-1},\label{Toddmassless}
\end{equation}
and the method to obtain the even-parity vector mode follows the process in Sec.~\ref{sec:EPV}. The isospectrality is illustrated in Fig.~\ref{ISOT}, and the corresponding effective potentials are provided in Fig.~\ref{ISOV}. In Fig.~\ref{ISOV}, it is clear that even in the massless limit of the even-parity vector mode, the effective potentials still $\omega$ dependent, and their maxima decrease as $\omega$ increases. In Fig.~\ref{ISOT}, Eq.~(\ref{Toddmassless}) is represented by the blue line, and the red dots represent the data points for even-parity vector modes in the massless limit. A selected set of values from Fig.~\ref{fig:isoevod} is listed in Table~\ref{Isisoectral}, where agreement to at least three decimal places is observed across the spectrum based on the third-order WKB results. This confirms the isospectrality between the odd-parity and even-parity vector modes in the massless limit. Therefore, although the effective potentials in Eqs.~(\ref{VSLRO}) and (\ref{VSLREV}) are explicitly different, the isospectrality of the greybody factors in the massless limit still holds.

\subsubsection{Comparison for the even-parity scalar and massive scalar modes}
\begin{figure}
    \centering
    \includegraphics[width=0.5\linewidth]{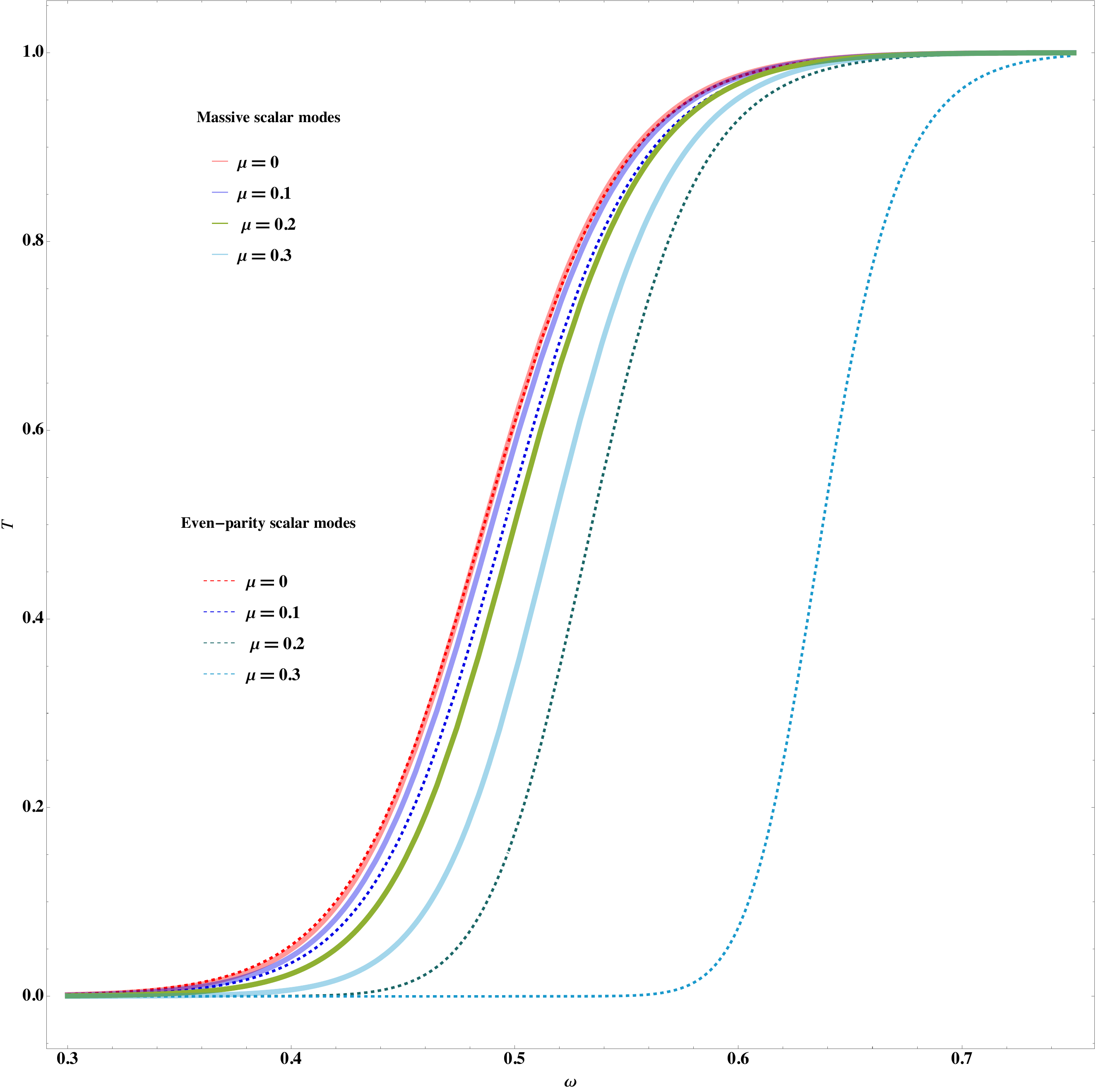}
    \caption{ The comparison for the greybody factor of the massive scalar perturbation and the even-parity scalar mode with $l = 2$ and varying $\mu$.}
    \label{fig:WKBesms}
\end{figure}
Since the leading-order term of the effective potential for even-parity scalar modes converges to the Regge–Wheeler scalar potential when the Proca mass is small, as shown in Eq.~(\ref{VSLRESep}), it is of interest to compare the greybody factors between even-parity scalar modes and massive scalar perturbations. This comparison is shown in Fig.~\ref{fig:WKBesms} for $l=2$, with varying $\mu = 0,~0.1,~0.2,~0.3$ for both cases. In the figure, the dotted lines represent even-parity scalar modes, while the solid lines correspond to massive scalar perturbations. In the massless limit, the even-parity vector mode behaves as a pure gauge mode, becoming degenerate with the massless scalar result. As the mass parameter $\mu$ increases, the greybody factors for both cases shift toward higher-energy regions. Consequently, the transmission probability becomes smaller than that in the massless case for a fixed $\omega$.

Moreover, the rightward shift of the greybody factor is consistently more significant for even-parity scalar modes than for massive scalar perturbations. This is consistent with the higher peak of the effective potential for even-parity scalar modes, as shown in Fig.~\ref{fig:effectivepotential}. The higher peak can also be attributed to the subleading term proportional to $\mu^2$ in Eq.~(\ref{VSLRESep}), where the coefficient is approximately $3f$ for even-parity scalar modes, in contrast to $f$ for massive scalar perturbations \cite{KONOPLYA2005377}. Therefore, for the same $l$, $\mu$, and $\omega$, the transmission probability of Proca particles with even-parity scalar modes is always lower than that of massive scalar particles, indicating that they are less likely to escape from the black hole.

\subsubsection{The turning behavior of the even-parity vector modes}
\begin{figure}
\begin{subfigure}{0.45\textwidth}
\includegraphics[width=\textwidth]{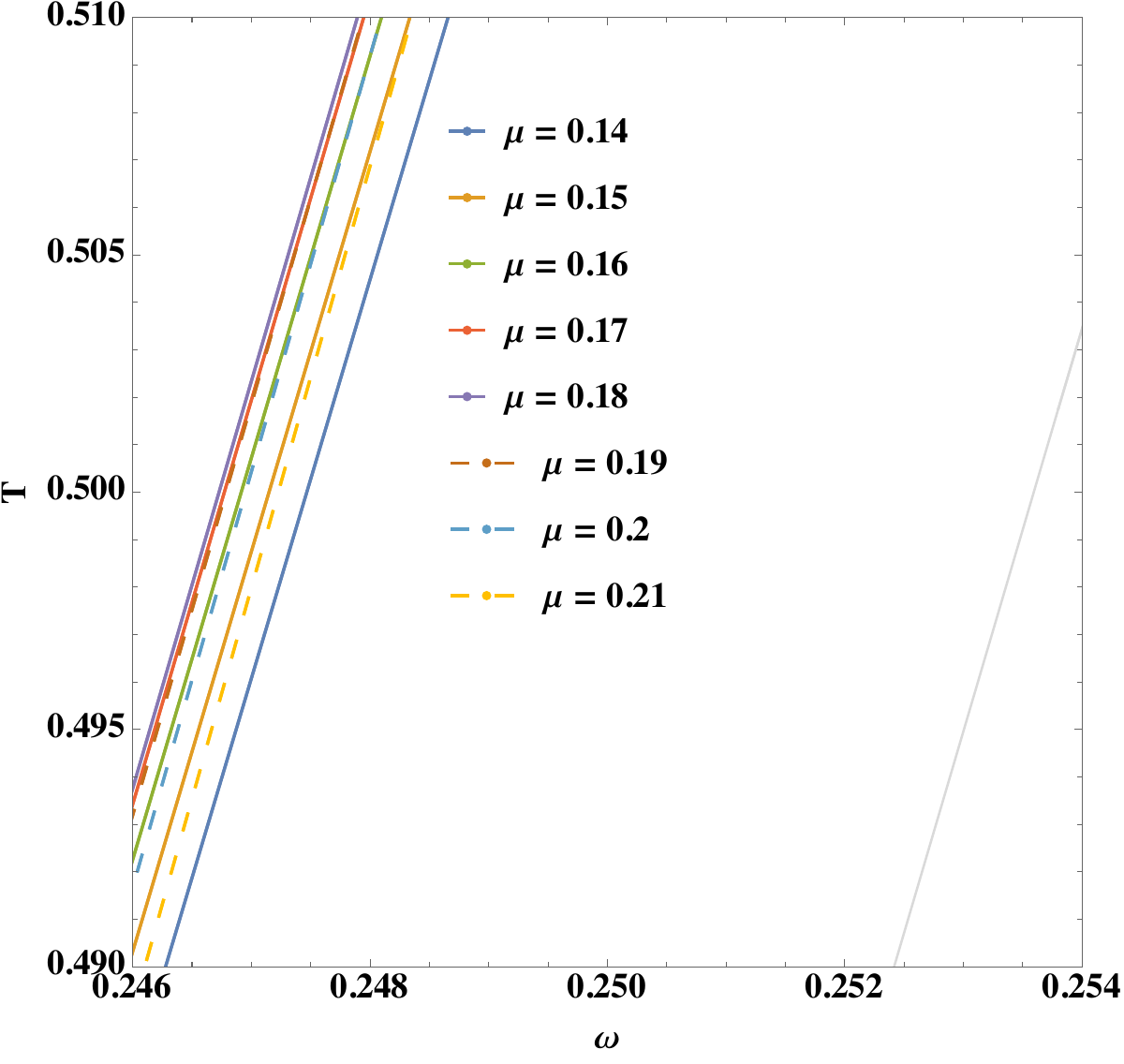}
\caption{$l=1$.}\label{turnCom1}
\end{subfigure}
\begin{subfigure}{0.435\textwidth}
\includegraphics[width=\textwidth]{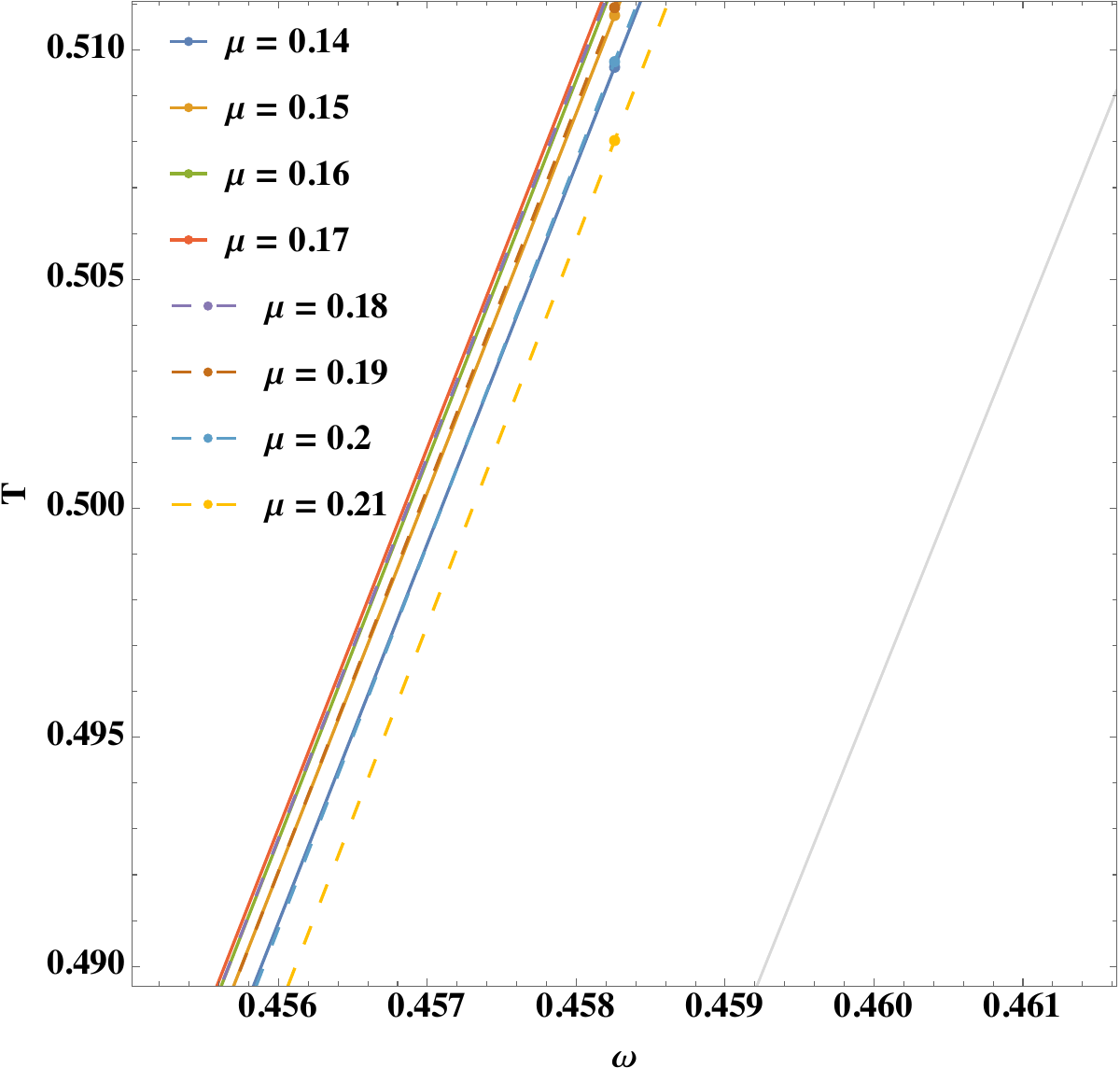}
\caption{$l=2$.}\label{turnCom2}
\end{subfigure}
\begin{subfigure}{0.45\textwidth}
\includegraphics[width=\textwidth]{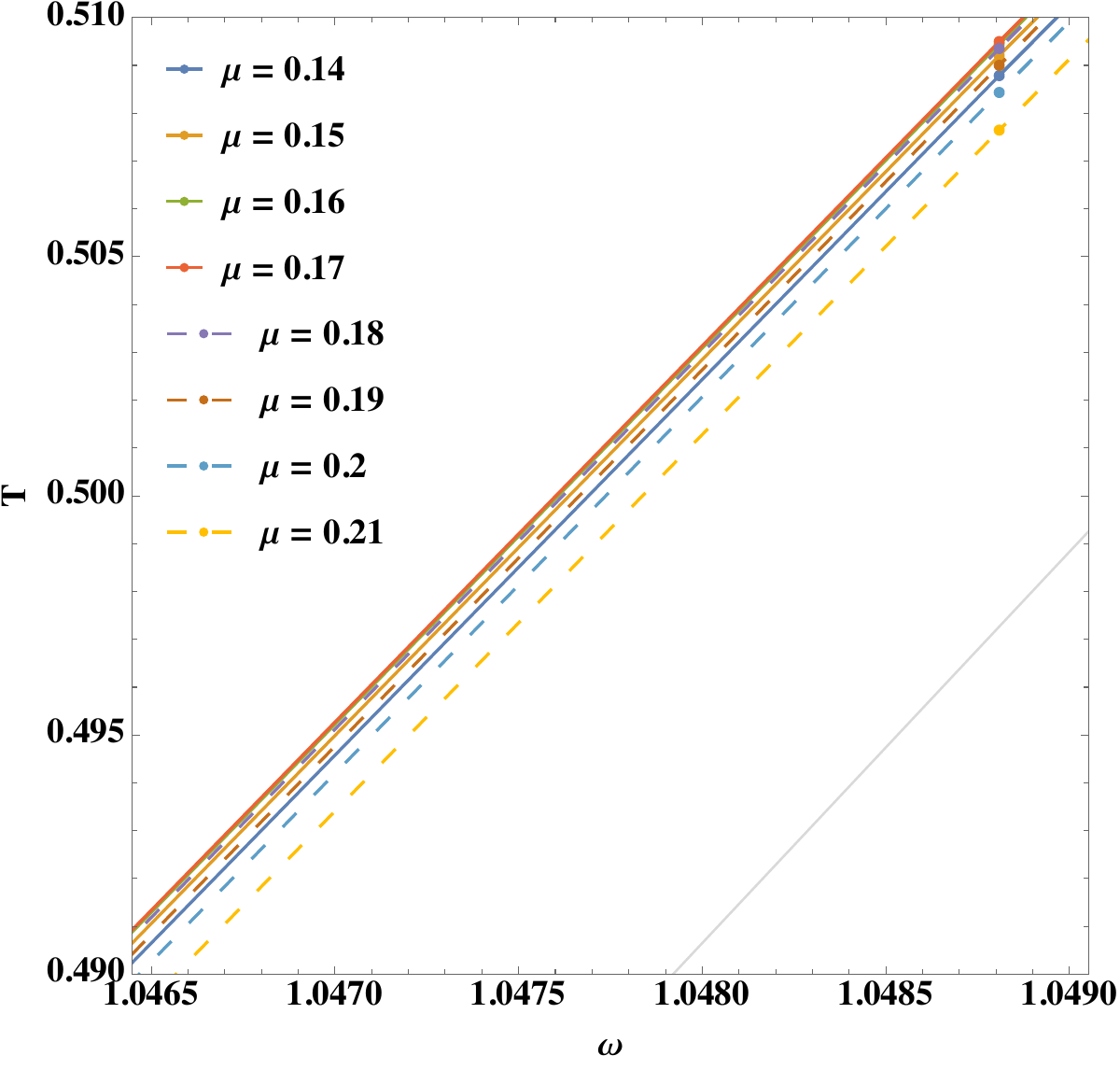}
\caption{$l=5$.}\label{turnCom5}
\end{subfigure}
\begin{subfigure}{0.45\textwidth}
\includegraphics[width=\textwidth]{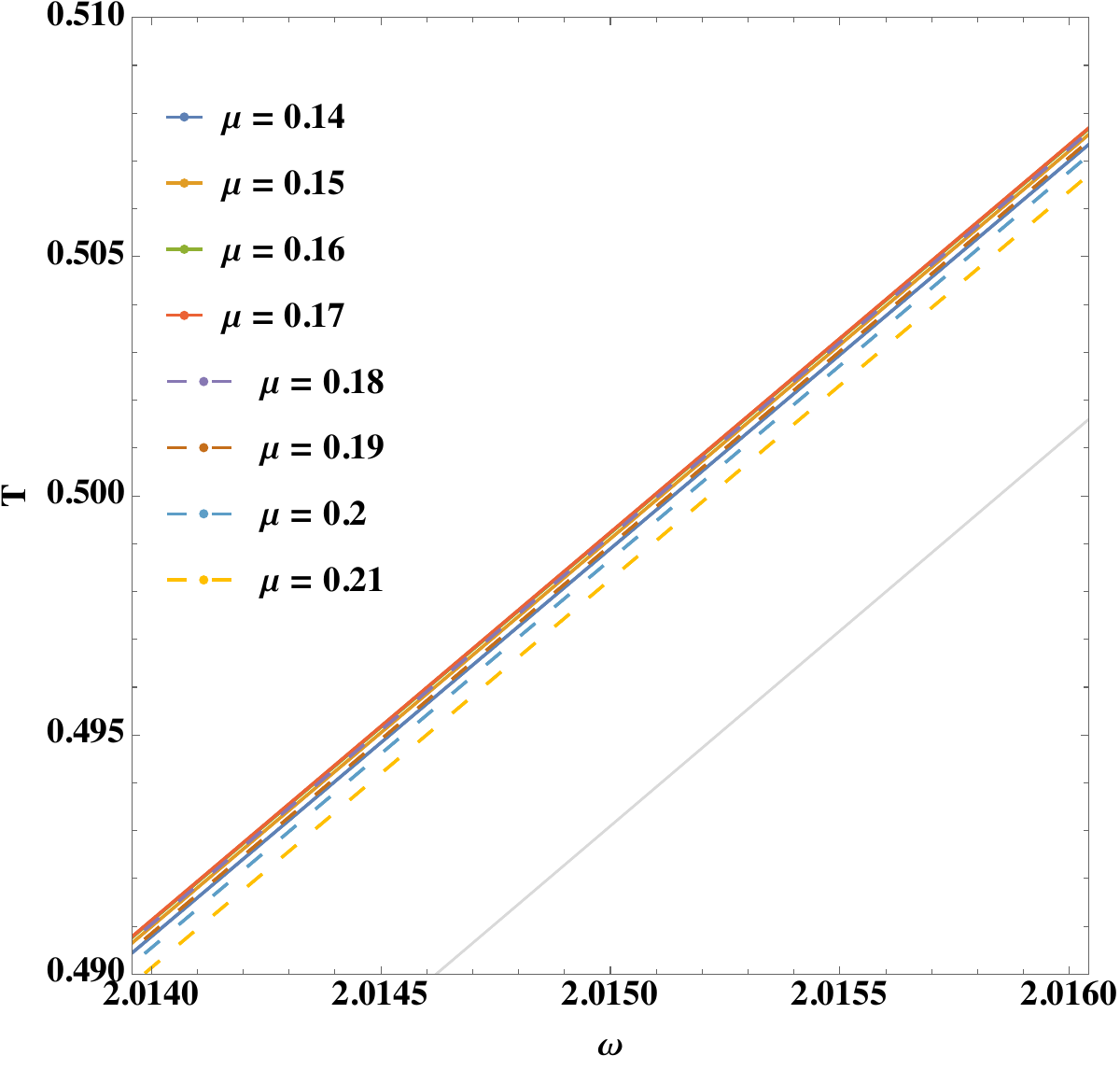}
\caption{$l=10$.}\label{turnCom10}
\end{subfigure}
\caption{The turning behavior for the even-parity vector modes with fixing the greybody factor in the region $T\approx0.5$. The gray line represented the massless results with the same $l$ in each sub-figure.}\label{turnCom}
\end{figure}
The turning behavior occurs only for the even-parity vector modes, which is supported by the effective potentials shown in Fig.~\ref{fig:effectivepotential}. More specifically, for a fixed set of $l$ and $\omega$, the maxima of the effective potentials first decrease and then increase as $\mu$ increases. In this region, the greybody factor for the Proca particle is larger than that for the photon with given specific values of $\omega$ and $l$. It is also true that for a fixed $l$, the Proca particle requires less energy than the photon to achieve the same transmission probability. In Fig.~\ref{turnCom}, we illustrate the turning behavior by varying $\mu$ in steps of $0.01$, within the range $0.14 < \mu < 0.21$, where the greybody factor is approximately $T \sim 0.5$. To clearly illustrate the turning behavior as $\mu$ increases, we use solid lines to represent left-shifting cases and dashed lines for right-shifting ones; the gray line corresponds to the massless vector case. In the following discussion, we analyze the behavior at $T \sim 0.5$, treating it as a representative slice of the parameter space.

A zoomed-in and more detailed version of the even-parity vector modes from Fig.~\ref{fig:WKB3modes} is shown in Fig.~\ref{turnCom2}, where the turning behavior occurs in the range $0.17 \lesssim \mu \lesssim 0.18$. We further examine the cases with $l=1,~5,10$ in Figs.~\ref{turnCom1}, ~\ref{turnCom5}, and ~\ref{turnCom10}. The turning behaviors around $T \sim 0.5$ occur in a similar region and do not explicitly depend on the variation of $l$. Nevertheless, the gap along the energy axis between the results and the massless case decreases as $l$ increases. This indicates that the difference in the required energy between the Proca particle and the photon to achieve the same greybody factor becomes less pronounced as $l$ increases.

\subsubsection{Example for the critical region of the even-parity vector modes}
\begin{figure}
\begin{subfigure}{0.45\textwidth}
\includegraphics[width=\textwidth]{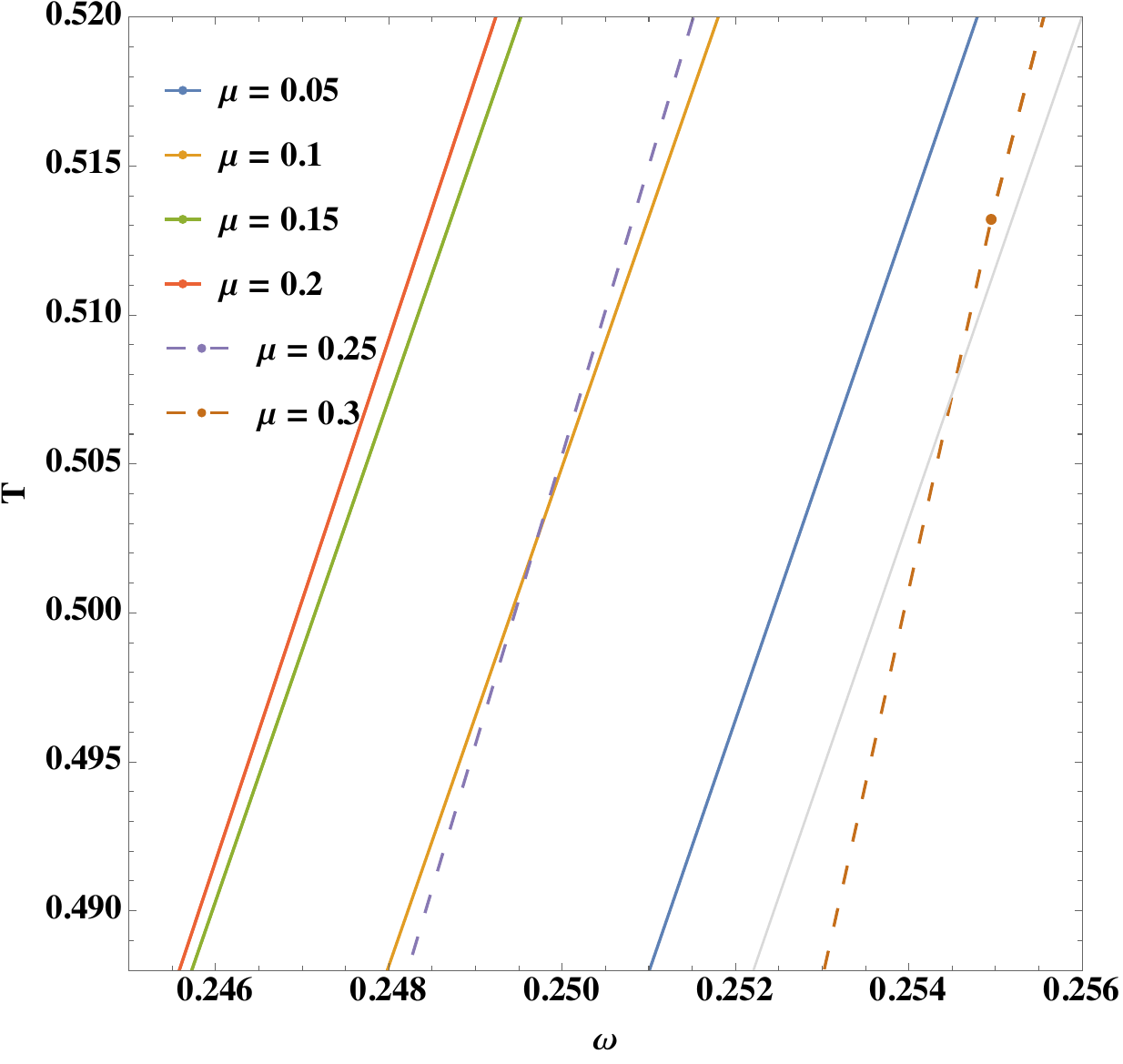}
\caption{$l=1$.}\label{turnCri1}
\end{subfigure}
\begin{subfigure}{0.43\textwidth}
\includegraphics[width=\textwidth]{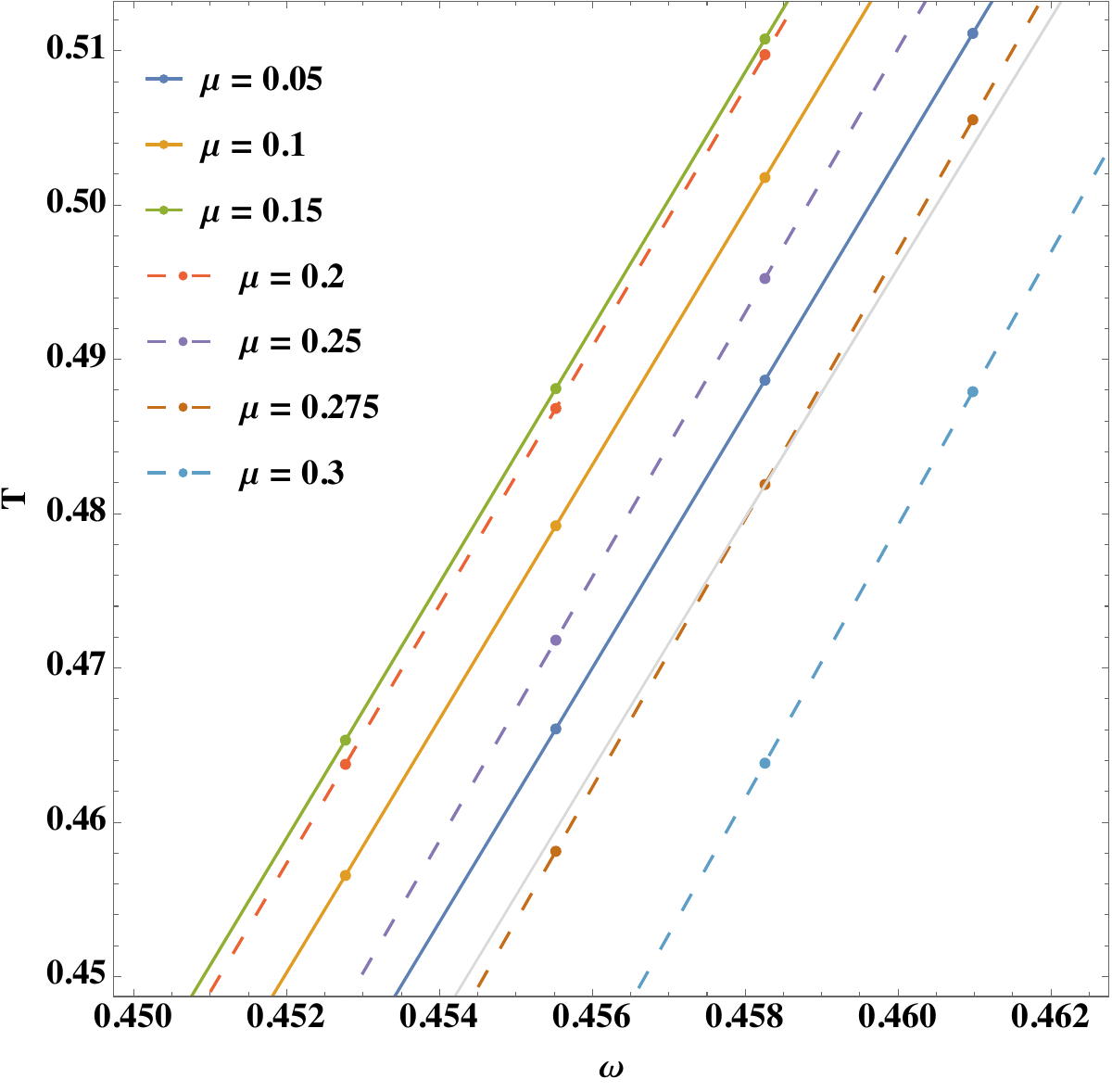}
\caption{$l=2$.}\label{turnCri2}
\end{subfigure}
\caption{The approximate critical value for the even-parity vector modes with fixing the greybody factor in the region $T\approx0.5$.  The gray line represented the massless results with the same $l$ in each sub-figure.}\label{truncri}
\end{figure}
After crossing the turning region, the greybody factor for even-parity vector modes exhibits a rightward shift as the Proca mass $\mu$ increases. Consequently, for each specific value of $l$ and transmission probability $T$, there exists a particular Proca mass at which the required energy $\omega$ matches that of the massless case. The set of all such parameter values is referred to as the ``critical value," which corresponds to a point on the diagram that overlaps with the massless curve. This critical point separates the regimes in which the Proca particle requires either less or more energy than the photon to achieve the same greybody factor.

In Fig.~\ref{truncri}, we present examples of critical values for $l = 1$ and $2$, where the gray line represents the massless vector curve. For $l = 1$, the critical point corresponding to the dashed and gray lines lies within the range $0.505 < T < 0.510$ at $\mu = 0.3$, while for $l = 2$, it appears around $T \sim 0.48$ at $\mu = 0.275$. The critical value of $\mu$ depends on $\omega$, $T$ and $l$. At the very least, we can confirm that $\mu = 0.3$ is not the critical value for the $l = 2$ case within the same region of $0.505 < T < 0.510$, as shown in Fig.~\ref{turnCri2}. Note that we follow the convention where solid lines represent left-shifting behavior, and dashed lines indicate right-shifting behavior. When comparing Fig.~\ref{turnCri1} with Fig.~\ref{turnCom1}, we see that the turning behavior occurs around $0.17 < \mu < 0.19$. Although $\mu = 0.2$ should be beyond the turning point, its curve still lies to the left of the $\mu = 0.15$ line in this region. Therefore, we continue using a solid line to represent it in Fig.~\ref{turnCri1}, consistent with the step size of $\mu = 0.05$ used in this analysis.

The behavior of the effective potential in Eq.~(\ref{VSLREV}) offers an explanation for the emergence of the critical value. In Figs.~\ref{fig:effectivepotentialfixlmu} and \ref{ISOV}, we present the typical decreasing behavior of the effective potentials as $\omega$ increases, for $\mu = 0.4$ and $0$, respectively. The decrease begins in the region where the peak of the effective potential is higher than that of the massless case. As $\omega$ increases, it crosses the massless curve and eventually falls below it. Physically, this implies that in the low-energy regime, the even-parity Proca particle experiences a stronger potential barrier than in the massless case, resulting in a lower transmission probability. In the high-energy regime, however, the Proca particle encounters a weaker barrier, leading to a higher transmission probability. Therefore, the critical value naturally emerges from this transition.

\begin{figure}
\begin{subfigure}{0.435\textwidth}
\includegraphics[width=\textwidth]{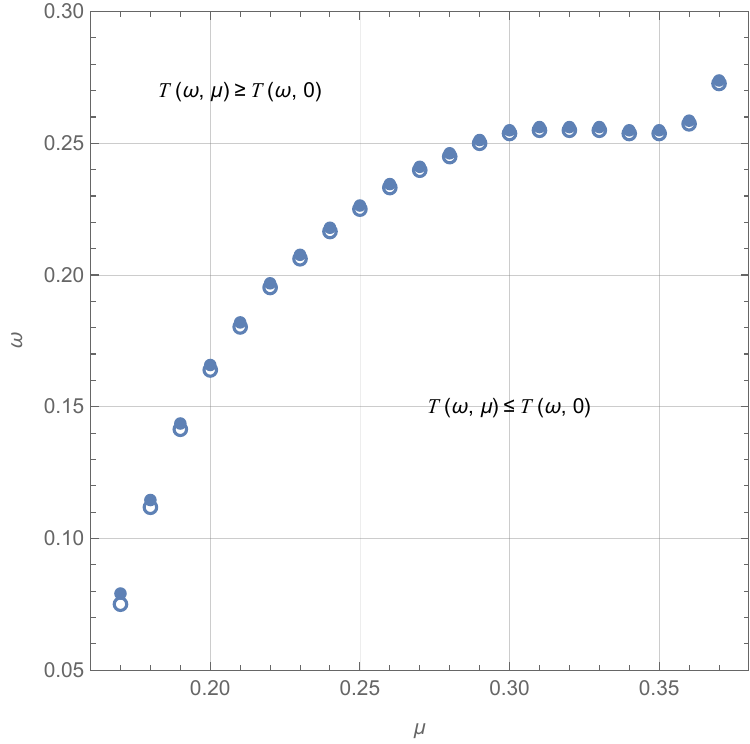}
\caption{$l=1$.}\label{CriPS1}
\end{subfigure}
\begin{subfigure}{0.42\textwidth}
\includegraphics[width=\textwidth]{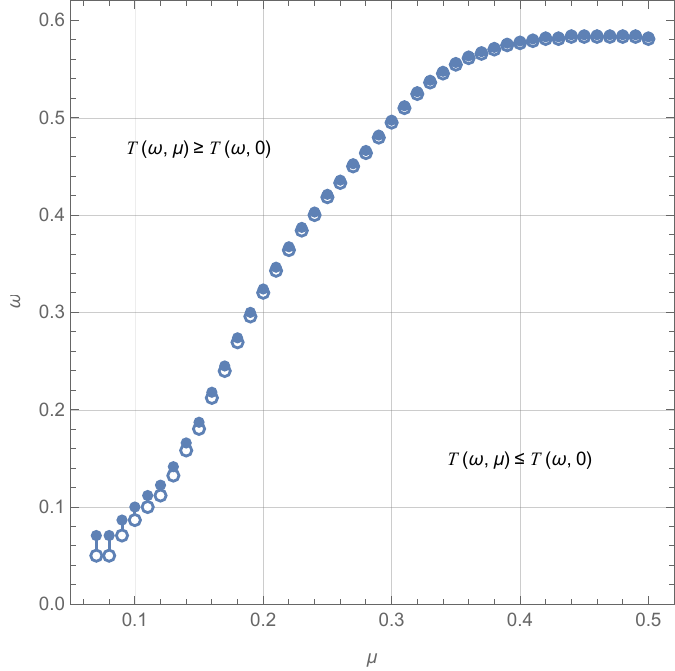}
\caption{$l=2$.}\label{CriPS2}
\end{subfigure}
\begin{subfigure}{0.435\textwidth}
\includegraphics[width=\textwidth]{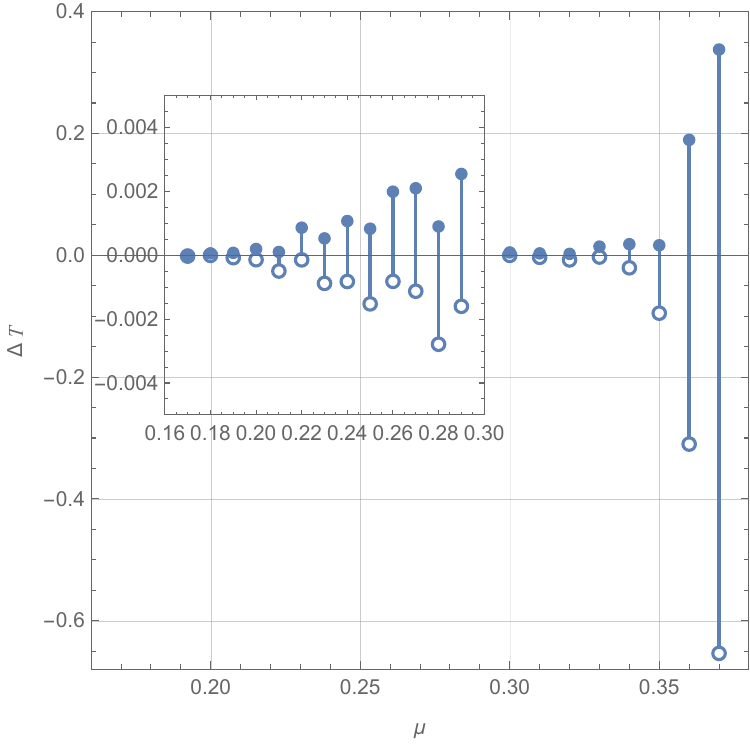}
\caption{$l=1$.}\label{CriDF1}
\end{subfigure}
\begin{subfigure}{0.45\textwidth}
\includegraphics[width=\textwidth]{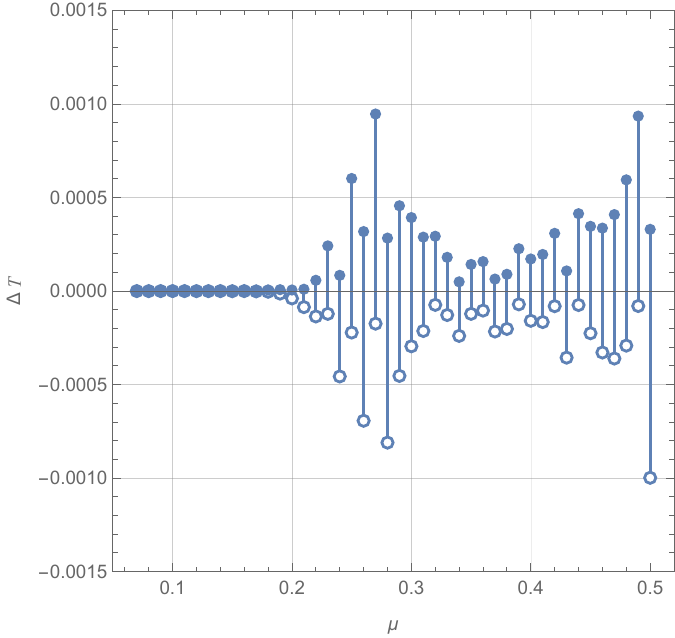}
\caption{$l=2$.}\label{CriDF2}
\end{subfigure}
\caption{The parameter space $\{\mu,\omega\}$ and the difference $\Delta T$ for the numerical results of the even-vector transmission probability that above/lower than the critical value when comparing with the massless vector one.}\label{CriCom}
\end{figure}

For a given set of $\{l,\,\mu\}$, a single critical point corresponding to a specific value of $\omega$ is expected for the even-parity vector modes. Nevertheless, the exact parameter set depends on the numerical accuracy. In our general approach, we set $\omega^{2}=0.005$ as the ``width'' of the numerical step, which means that we compute ten data points in the range $0.2\leq \omega <0.3$, and more or fewer points in the ranges $0.3\leq \omega <0.4$ and $0.1\leq \omega <0.2$, etc. By comparing the corresponding greybody factors with those of the massless even-parity vector mode, we define the difference as $\Delta T = T(\omega,\mu)-T(\omega,0)$ for all data points, select the neighboring points for which $\Delta T$ exhibits a sign change from ``$-$'' to ``$+$'', and confine the exact critical point between the corresponding $\omega$ values of this neighborhood. In this way, $|\Delta T|$ serves as an ``error estimate'' for the exact critical point, and one can always decrease the width of the numerical step to constrain a narrower region that corresponds to a smaller $|\Delta T|$. In Sec.~\ref{sec:Iso}, the isospectrality of the greybody factors for the massless odd- and even-parity vector modes holds up to the order of $10^{-3}$, which we may regard as the effective accuracy of the current approach. Therefore, decreasing the width of $\omega^{2}$ until $|\Delta T|\sim 10^{-3}$ should be sufficient for efficiently analyzing the critical points.
 
 The confining region of the critical points for the selected parameter sets $\{\mu,\omega\}$ with $l=1,~2$ are shown in Figs.~\ref{CriPS1} and~\ref{CriPS2}, where the ``dot'' represents the first data point at which $\Delta T > 0$ for each $\mu$, and the ``circle'' represents the neighboring point at which $\Delta T<0$. The widths of the numerical steps are given by $\omega^{2}=6.25\times 10^{-4}$ for the $l=1$ cases and $\omega^{2}=2.5\times 10^{-3}$ for the $l=2$ case. Accordingly, in Figs.~\ref{CriPS1} and~\ref{CriPS2}, the exact value of $\omega$ for each ``dot'' is given by $\omega=\sqrt{N\times6.25\times10^{-4}}$ and $\omega=\sqrt{N\times2.5\times10^{-3}}$, respectively, with a positive integer $N$, while the corresponding values for each ``circle'' are given by $\omega=\sqrt{(N-1)\times6.25\times10^{-4}}$ and $\omega=\sqrt{(N-1)\times2.5\times10^{-3}}$. The exact parameter set corresponding to the critical point in each case is thus confined to the interval between the corresponding ``dot'' and ``circle''. The corresponding values of $\Delta T$ for the ``dot'' and ``circle'' in Figs.~\ref{CriPS1} and~\ref{CriPS2} are shown in Figs.~\ref{CriDF1} and~\ref{CriDF2}, respectively, with the same symbols used to denote the ``dot'' and ``circle''.
 
 For the $l=1$ cases, the focusing region for massless vector particles is $0.12\lesssim \omega \lesssim 0.40$, with transmission probabilities in the range $0.01\lesssim T\lesssim 0.99$. From Fig.~\ref{CriPS1}, one may expect that the full spectra for $\mu < 0.18$ exhibit higher transmission probabilities than those for photons. For $0.18\leqslant \mu \leqslant 0.25$, the critical points occur at nonvanishing greybody factors, causing the transmission probabilities to split into values either lower or higher than those of photons, depending on the critical parameter set $\{\omega,\mu\}$. Note that $\Delta T$ for $\mu\leqslant0.25$ is sufficiently minimized to the order of $\sim10^{-3}$ with the width setting $\omega^{2}=6.25\times10^{-4}$, as shown in the subfigure of Fig.~\ref{CriDF1}. For $\mu>0.25$, the difference bars in Fig.~\ref{CriDF1} begin to increase, particularly for $\mu>0.36$, indicating a slight enhancement of numerical uncertainty within the selected $\omega$ range. The physical interpretation is twofold: the effective potentials approach the barrier-like limit for these parameter sets, and $\omega^{2}\lesssim V_{eff}^{(\text{even-vector})}|_{r\rightarrow\infty}=\mu^{2}$ in this region. Note that the greybody factor for the $l=1$ massless vector particle approaches unity as $\omega\rightarrow 0.52$. Consequently in Fig.~\ref{CriPS1}, for nonvanishing greybody factors, the equality condition in the region $T(\omega,\mu)\geqslant T(\omega,0)$ occurs around $\omega\gtrsim0.52$, where the massless results converge to unity. In contrast, in the region $T(\omega,\mu)\leqslant T(\omega,0)$, the equality occurs when the transmission probability of the massive particle converges to unity.

For the $l=2$ cases, the focusing region for the massless vector particles is $0.32\lesssim \omega \lesssim 0.60$, with transmission probabilities in the range $0.01\lesssim T\lesssim 0.99$. Therefore, the full spectra for $\mu \leqslant 0.19$ exhibit higher transmission probabilities than those of photons. For $0.19<\mu\leqslant 0.5$, we have obtained the critical spectrum within the range $|\Delta T|\lesssim 10^{-3}$, which distinguishes whether the transmission probabilities are higher or lower than those of the $l=2$ massless vector particle, as illustrated in Figs.~\ref{CriPS2} and~\ref{CriDF2}. The equality condition for the region $T(\omega,\mu)\geqslant T(\omega,0)$ appears around $\omega\gtrsim 0.92$, where the greybody factors of the $l=2$ massless vector particle approach unity. The effective potential becomes non–barrier-like, approximately around $\mu\gtrsim 0.6$ for the $l=2$ cases, therefore, a behavior of increasing $\Delta T$ similar to that in the $l=1$ case is expected to occur in this parameter space.

\subsubsection{Extract the critical point with secant method}\label{sec:critical secant}
With the critical region already confined, we may further extract a single critical point for each parameter set by applying the secant method \cite{atkinson1989numerical}, which is a well-defined iterative process for finding the root of a function without requiring the first-order derivative in numerical analysis. The root of the function $\Delta T(\omega)=0$, where $\Delta T(\omega)=T(\omega,\mu)-T(\omega,0)$ has the same definition as in the previous subsection, can be obtained by constructing a line through the points $(\omega_{0},\ \Delta T(\omega_{0}))$ and $(\omega_{1},\ \Delta T(\omega_{1}))$, and solving the critical point $\omega_{i\geq2}$ by iterating
\begin{equation}\label{secant method}
\omega_{2}=\omega_{1}-\Delta T(\omega_{1})\frac{\omega_{1}-\omega_{0}}{\Delta T(\omega_{1})-\Delta T(\omega_{0})}.
\end{equation} 

For a given $\mu$, the neighboring points that confine the critical region in the previous subsection can be used as the initial parameters, $\omega_{0}$ and $\omega_{1}$, in Eq.~(\ref{secant method}). Since the initial parameters already confine a small region, the resulting value $\omega_{2}$ obtained from Eq.~(\ref{secant method}) typically leads to convergence with $|\Delta T(\omega_{2})| \lesssim 10^{-5}$, except in cases where the effective potential becomes non-barrier-like. Therefore, the parameter set $(\mu, \omega_{2})$ for sufficiently convergent cases can be treated as critical points. The corresponding critical points related to Fig.~\ref{CriCom} are presented in Tables~\ref{Tab:criticalpsl1} and~\ref{Tab:criticalpsl2}.
 
\begin{table}[t!]
\centering
\caption{The critical points $(\mu,\omega_{2})$, the corresponding greybody factors $T(\omega_{2})$, and $|\Delta T(\omega_{2})|$ for the even-parity vector modes with $l=1$.}\label{Tab:criticalpsl1}
\begin{tabular}{cccccccc}
\hline
                & $\mu=0.17$ & $\mu=0.18$ & $\mu=0.19$ & $\mu=0.20$ & $\mu=0.21$ & $\mu=0.22$ & $\mu=0.23$  \\
\hline
$\omega_{2}$    & $0.077611$ & $0.112044$ & $0.142523$ & $0.164723$ & $0.181708$ & $0.195485$ & $0.207099$ \\
$T(\omega_{2})$ & $0.002976$ & $0.007722$ & $0.020280$ & $0.042257$ & $0.073925$ & $0.114798$ & $0.163457$ \\
$|\Delta T(\omega_{2})|$ & $6.24\times10^{-7}$ & $3.83\times10^{-7}$ & $2.67\times10^{-6}$ & $4.83\times10^{-6}$ & $4.24\times10^{-6}$ & $4.73\times10^{-6}$ & $9.92\times10^{-6}$ \\
\hline
                & $\mu=0.24$ & $\mu=0.25$ & $\mu=0.26$ & $\mu=0.27$ & $\mu=0.28$ &  $\mu=0.29$ & $\mu=0.30$  \\
\hline
$\omega_{2}$    & $0.217131$ & $0.225896$ & $0.233574$ & $0.240246$ & $0.245911$ & $0.250481$ & $0.253769$ \\
$T(\omega_{2})$ & $0.217683$ & $0.274578$ & $0.331079$ & $0.384265$ & $0.431499$ & $0.470396$ & $0.498492$ \\
$|\Delta T(\omega_{2})|$ & $9.49\times10^{-6}$ & $6.35\times10^{-6}$ & $1.52\times10^{-6}$ & $4.04\times10^{-6}$ & $8.62\times10^{-6}$ & $2.05\times10^{-5}$ & $5.04\times10^{-6}$ \\
\hline
                & $\mu=0.31$ & $\mu=0.32$ & $\mu=0.33$ & $\mu=0.34$ & $\mu=0.35$ &  $\mu=0.36$ & $\mu=0.37$  \\
\hline
$\omega_{2}$    & $0.255587$ & $0.255889$ & $0.255158$ & $0.254374$ & $0.254770$ & $0.258142$ & $0.273472$ \\
$T(\omega_{2})$ & $0.514056$ & $0.516636$ & $0.510436$ & $0.503945$ & $0.507215$ & $0.535133$ & $1.000000$ \\
$|\Delta T(\omega_{2})|$ & $5.27\times10^{-5}$ & $6.07\times10^{-5}$ & $8.71\times10^{-5}$ & $2.93\times10^{-4}$ & $1.80\times10^{-4}$ & $5.77\times10^{-4}$ &
$0.340733$ \\
\hline
\end{tabular}
\end{table}

\begin{table}[t!]
\centering
\caption{The critical points $(\mu,\omega_{2})$, the corresponding greybody factors $T(\omega_{2})$, and $|\Delta T(\omega_{2})|$ for the even-parity vector modes with $l=2$.}\label{Tab:criticalpsl2}
\begin{tabular}{cccccccc}
\hline
                & $\mu=0.07$ & $\mu=0.08$ & $\mu=0.09$ & $\mu=0.10$ & $\mu=0.11$ & $\mu=0.12$ & $\mu=0.13$  \\
\hline
$\omega_{2}$    & $0.055848$ & $0.066165$ & $0.077322$ & $0.090032$ & $0.104138$ & $0.120391$ & $0.138841$ \\
$T(\omega_{2})$ & $0.000011$ & $0.000012$ & $0.000014$ & $0.000017$ & $0.000021$ & $0.000028$ & $0.000041$ \\
$|\Delta T(\omega_{2})|$ & $8.60\times10^{-9}$ & $8.79\times10^{-9}$ & $1.34\times10^{-8}$ & $1.15\times10^{-8}$ & $1.53\times10^{-8}$ & $1.18\times10^{-8}$ & $1.82003\times10^-8$ \\
\hline
                & $\mu=0.14$ & $\mu=0.15$ & $\mu=0.16$ & $\mu=0.17$ & $\mu=0.18$ &  $\mu=0.19$ & $\mu=0.20$  \\
\hline
$\omega_{2}$    & $0.160220$ & $0.184772$ & $0.212476$ & $0.241717$ & $0.271005$ & $0.298506$ & $0.323535$ \\
$T(\omega_{2})$ & $0.000066$ & $0.000123$ & $0.000265$ & $0.000637$ & $0.001618$ & $0.004014$ & $0.009327$ \\
$|\Delta T(\omega_{2})|$ & $2.41\times10^{-8}$ & $3.83\times10^{-8}$ & $1.74\times10^{-8}$ & $1.39\times10^{-7}$ & $3.04\times10^{-7}$ & $6.38\times10^{-7}$ & $6.47\times10^{-7}$ \\
\hline
                & $\mu=0.21$ & $\mu=0.22$ & $\mu=0.23$ & $\mu=0.24$ & $\mu=0.25$ &  $\mu=0.26$ & $\mu=0.27$  \\
\hline
$\omega_{2}$    & $0.346043$ & $0.366406$ & $0.385144$ & $0.402628$ & $0.419132$ & $0.434985$ & $0.450430$ \\
$T(\omega_{2})$ & $0.019968$ & $0.039461$ & $0.072602$ & $0.124752$ & $0.200010$ & $0.299316$ & $0.417467$ \\
$|\Delta T(\omega_{2})|$ & $1.01\times10^{-6}$ & $4.08\times10^{-6}$ & $6.72\times10^{-6}$ & $5.01\times10^{-6}$ & $7.44\times10^{-6}$ & $5.45\times10^{-6}$ & $3.31\times10^{-7}$ \\
\hline
                & $\mu=0.28$ & $\mu=0.29$ & $\mu=0.30$ & $\mu=0.31$ & $\mu=0.32$ &  $\mu=0.33$ & $\mu=0.34$  \\
\hline
$\omega_{2}$    & $0.465673$ & $0.480879$ & $0.496056$ & $0.510942$ & $0.524883$ & $0.537156$ & $0.547333$ \\
$T(\omega_{2})$ & $0.543090$ & $0.662180$ & $0.762794$ & $0.838922$ & $0.890859$ & $0.923741$ & $0.943836$ \\
$|\Delta T(\omega_{2})|$ & $4.88\times10^{-6}$ & $9.99\times10^{-6}$ & $9.70\times10^{-6}$ & $7.77\times10^{-6}$ & $3.76\times10^{-6}$ & $4.46\times10^{-6}$ &$2.25\times10^{-6}$ \\
\hline
                & $\mu=0.35$ & $\mu=0.36$ & $\mu=0.37$ & $\mu=0.38$ & $\mu=0.39$ &  $\mu=0.40$ & $\mu=0.41$  \\
\hline
$\omega_{2}$    & $0.555565$ & $0.562136$ & $0.567387$ & $0.571605$ & $0.574975$ & $0.577669$ & $0.579780$ \\
$T(\omega_{2})$ & $0.956359$ & $0.964417$ & $0.969824$ & $0.973596$ & $0.976284$ & $0.978245$ & $0.979674$ \\
$|\Delta T(\omega_{2})|$ & $3.52\times10^{-6}$ & $3.30\times10^{-6}$ & $2.52\times10^{-6}$ & $3.17\times10^{-6}$ & $2.83\times10^{-6}$ & $4.30\times10^{-6}$ & $4.73\times10^{-6}$ \\
\hline
                & $\mu=0.42$ & $\mu=0.43$ & $\mu=0.44$ & $\mu=0.45$ & $\mu=0.46$ &  $\mu=0.47$ & $\mu=0.48$  \\
\hline
$\omega_{2}$    & $0.581393$ & $0.582597$ & $0.583422$ & $0.583940$ & $0.584152$ & $0.584098$ & $0.583799$ \\
$T(\omega_{2})$ & $0.980703$ & $0.981441$ & $0.981930$ & $0.982235$ & $0.982359$ & $0.982330$ & $0.982158$ \\
$|\Delta T(\omega_{2})|$ & $3.48\times10^{-6}$ & $4.40\times10^{-6}$ & $3.57\times10^{-6}$ & $7.75\times10^{-6}$ & $9.59\times10^{-6}$ & $1.14\times10^{-5}$ & $1.21\times10^{-5}$ \\
\hline
\end{tabular}
\end{table}
In summary, when we compare the greybody factor of Proca particles with that of a photon for the same $l$, critical points can occur for all allowed values of $\mu$ constrained by the barrier-like potential. These critical points classify the resulting greybody factors into three distinct regions that are higher than, coincident with, or lower than the photon result, as shown in Fig.~\ref{CriCom}.  For a given set of $l$ and $\mu$, if the critical value occurs at small $\omega$ with $T\sim0$, as in the first row of Table~\ref{Tab:criticalpsl1}, the first two rows of Table~\ref{Tab:criticalpsl2}, and for values of $\mu$ smaller than those cases, the greybody spectrum within $0<T<1$ is effectively shifted to the left relative to the massless vector case, which may yield a higher transmission probability than that of the photon for fixed $\omega$. If the critical value lies within the range $0 < T < 1$ for a given $\mu$, it defines a boundary in the diagram that separates regions with higher and lower transmission probabilities than the photon, with the critical $\omega$ serving as a reference cut, as shown in Figs.~\ref{truncri} and~\ref{CriCom}, the second and third rows of Table~\ref{Tab:criticalpsl1}, and the third and fourth rows of Table~\ref{Tab:criticalpsl2}. Lastly, if the critical value occurs around $T \approx 1$ for a given $\mu$, as in the last two rows of Table~\ref{Tab:criticalpsl2}, the entire spectrum lies to the right of the massless vector case, implying that the transmission probability is always lower than that of the photon for the same $\omega$.

\subsubsection{Discussion}\label{sec:compareref}
The comparison between our results and the results in \cite{HerSamWan2012} is presented in this subsection. In their work, the three modes for the Proca fields were denoted as $l_{T}$, $l_{1}$, and $l_{2}$, and the massless results were denoted as $l_{E}$. The $l_{1}$ and $l_{2}$ modes were evaluated from the coupled radial equations, which correspond to the even-parity modes in our work, while the $l_{T}$ mode corresponds to the odd-parity mode evaluated by the decoupled radial equation. The $l_{T}$ and $l_{1}$ modes degenerate with the $l_{E}$ mode in the massless limit; therefore, the $l_{1}$ mode is our even-parity vector mode. The $l_{2}$ mode corresponds to the even-parity scalar mode, which is absent in Maxwell theory but exists for non-exactly vanishing Proca mass, as also mentioned in their description. In Fig.~1 of \cite{HerSamWan2012}, for a fixed $l$, the relative ordering of the transmission probabilities of these modes is $l_{1}$, $l_{T}$, and $l_{2}$ from the lower to higher $\omega r_{H}$ region for non-vanishing Proca mass in the four-dimensional Schwarzschild background, which is consistent with our ordering results presented in Figs.~\ref{fig:sbmu} and \ref{fig:WKB3modes}. The relative shifting for various $l$ is also consistent with Fig.~\ref{fig:sbl}.

The scale differs when comparing our results with those in \cite{HerSamWan2012} due to the choice of units. We set the black hole mass as the unit, while the reference adopts the event horizon radius as the unit, $r_{H}=1$. Accordingly, for our massless results shown in Fig.~\ref{fig:isoevod}, the results become consistent if we rescale the $x$-axis as $\omega r_{H}=2\omega$. A particular point noted in \cite{HerSamWan2012} is that for $l_{1}=1$ (corresponding to our even-parity vector modes) and a Proca mass of $\mu r_{H}=0.3$, the transmission probability starts at a nonzero frequency. We do not observe the same phenomenon with the corresponding set of parameters, which correspond to $\mu=0.15$ and $r_{H}=2$ in this work. We present selected results for the even-parity vector modes with $\mu=0.15$ in the first two rows of Table~\ref{Tab:lex}. The results in the first row belong to the absent region in the reference, where $\omega r_{H}<\mu r_{H}$. The first entry of the second row corresponds exactly to the nonvanishing starting point reported in the reference, for which the transmission probabilities are consistent. The remaining results are also consistent with those in the reference after rescaling $\omega$ by $r_{H}$. In summary, the WKB approximation allows us to compute the greybody factors for cases with $\omega r_{H}<\mu r_{H}$ and yields nonvanishing results; on the other hand, the results for the $\omega r_{H}>\mu r_{H}$ cases are generally consistent with those reported in the reference.

\begin{figure}
    \centering
    \includegraphics[width=0.5\linewidth]{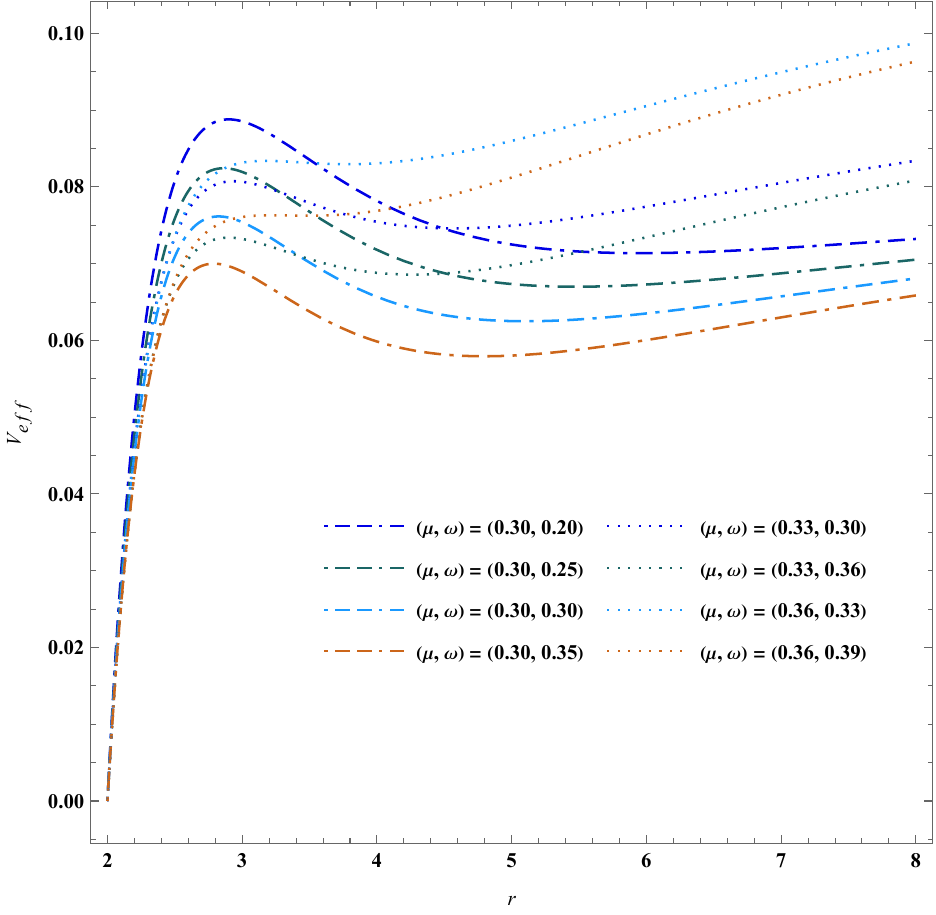}
    \caption{ The effective potentials for the even-parity vector modes with $l=1$ and selected $(\mu,\omega)$.}
    \label{fig:effectivepotentiall1ex}
\end{figure}

For the cases of the even-parity vector modes with $l=1$ and $\mu\geq 0.33$, the greybody factors include irregular increases and decreases as $\omega$ increases, and for $\mu\geq0.36$ there appears a gap in which the results increase immediately from nearly vanishing to nearly unity within a small range of $\omega$ as presented in the last two rows of Table~\ref{Tab:lex}. The latter phenomenon can be regarded as analogous to the transmission probability starting from nonzero $\omega$ and $T\sim 1$ for the case of $l_{1}=1$ and $\mu r_{H}=1$ in the reference, where the mass parameter corresponds to $\mu=0.5$ in our case and satisfies the condition $\mu\geq0.36$. The physical explanation shall refer to the non-barrier-like effective potentials. Note that the non-barrier-like effective potential includes two cases: one in which the peak is lower than the asymptotic behavior of the effective potential, and one resembling a step-function-like potential. Both types of effective potentials occur naturally for the even-parity vector modes as $\omega$ increases, as discussed in Ch.~\ref{sec:effp}. Even though the accuracy of the WKB method for nearly non-barrier-like potentials involves uncertainties, the selected effective potentials and the corresponding greybody factors discussed in this paragraph are presented in Fig.~\ref{fig:effectivepotentiall1ex} and Table~\ref{Tab:lex}. As a remark, for the cases with $l_{T}=1$ and $l_{2}=1$ and $\mu r_{H}=1$ in the reference, although the related modes are odd- and even-parity scalar modes, the corresponding effective potentials are clearly non-barrier-like; therefore, the nonzero starting values of the greybody factors in these two cases can be classified into the category described in this paragraph.
\begin{table}[t!]
\centering
\caption{The greybody factors for the even-parity vector modes with $l=1$ and selected $(\mu,\omega)$.}\label{Tab:lex}
\begin{tabular}{cccccccc}
\hline
    & &  & & $(\omega,T)$ &  &  &   \\
\hline
$\mu=0.15$ & $(0.025, 0.001)$ & $(0.056, 0.002)$ & $(0.075, 0.003)$ & $(0.103, 0.007)$ & $(0.117, 0.011)$ & $(0.125, 0.014)$ & $(0.141, 0.024)$ \\
           & $(0.150, 0.032)$ & $(0.200, 0.162)$ & $(0.225, 0.318)$ & $(0.250, 0.524)$ & $(0.275, 0.717)$ & $(0.300, 0.850)$ &
$(0.325, 0.927)$ \\
\hline
$\mu=0.30$ & $(0.212, 0.061)$ & $(0.235, 0.247)$ & $(0.255, 0.513)$ & $(0.274, 0.702)$ & $(0.292, 0.805)$ & $(0.308, 0.867)$ & $(0.324, 0.917)$ \\
$\mu=0.33$ & $(0.235, 0.077)$ & $(0.255, 0.506)$ & $(0.274, 0.771)$ & $(0.292, 0.807)$ & $(0.308, 0.778)$ & $(0.324, 0.828)$ & $(0.346, 0.997)$ \\
$\mu=0.36$ & $(0.255, 0.002)$ & $(0.274, 1.000)$ & $(0.292, 0.999)$ & $(0.308, 0.000)$ & $(0.324, 0.000)$ & $(0.346, 1.000)$ & $(0.367, 1.000)$ \\
\hline
\end{tabular}
\end{table}

\section{Conclusion}\label{sec:conclu}
We study the greybody factor for the Proca field in Schwarzschild black hole spacetime. The separated radial equations are obtained using the VSH method, and the even-parity equations are further decoupled through a set of transformations corresponding to the FKKS separation in the static limit. We analyze that the separation constant in the FKKS ansatz under the static limit can be reproduced by applying the Lorenz condition as an alternative and equivalent formulation, and we rearrange the separated and decoupled radial equations into a Schrödinger-like form. The radial equations in this Schrödinger-like form allow us to study the behavior of the effective potentials and serve as a foundation for interpreting the greybody factor results.

The radial equations for the Proca field contain three degrees of freedom: the even-parity vector and odd-parity modes correspond to vector-type polarizations, while the even-parity scalar mode corresponds to scalar-type polarization. The greybody factors for these modes are studied using two semi-analytical methods: the rigorous bound and the WKB approximation. The results from both methods are generally consistent. The rigorous bound provides a lower limit, while the WKB approximation improves accuracy. For a unified set of parameters, the spectral ordering is the same in both methods: the even-parity vector mode appears at the lowest energy, followed by the odd-parity mode, and the even-parity scalar mode at the highest energy. This ordering remains valid when the Proca mass $\mu$ is fixed and the angular momentum number $l$ is varied, with all modes shifting toward higher energy as $l$ increases.

The even-parity scalar modes behave as the Regge-Wheeler scalar equation in the massless limit. Even though the scalar-type polarization behaves as pure gauge modes in Maxwell theory within the Schwarzschild black hole geometry, the comparison of Proca perturbations with the massive scalar case is achieved in this study. For a given set of common parameters, the massive scalar particle exhibits a higher transmission probability than the even-parity scalar modes in the Schwarzschild background. The spectra shift toward higher energy regions compared to the massless scalar case as the particle mass increases, which is consistent with the general expectation that massive particles have lower transmission probabilities than massless ones. However, for a given set of common parameters, the vector particle exhibits a higher transmission probability than the scalar particle in the massless limit. In contrast, in the massive case, a specific mode, the even parity scalar mode, exhibits behavior opposite to that observed in the massless case.

The degeneracy of the vector-type polarization in the massless limit cannot be directly observed from the radial equations of the even-parity vector and odd-parity modes in their Schrödinger-like forms. The greybody factors obtained through the rigorous bound are nearly isospectral, though still distinguishable; however, this isospectrality is confirmed by the WKB results up to three decimal places. Therefore, the degeneracy in the physical behavior of the vector-type polarization holds in the massless limit. However, the effective potentials take different forms in the massless limit; the explicit difference lies in the $\omega$ dependence of the even-parity vector modes, whereas the odd-parity modes are independent of $\omega$. Whether there exists a Chandrasekhar-like transformation similar to those relating the Regge–Wheeler and Zerilli equations \cite{ChandraDetw1975}, which leads to a supersymmetric partner potential as discussed in \cite{PhysRevD.62.064009}, remains an open question.

With other parameters fixed and the mass parameter set to be massive, the greybody factor of the vector-type polarization splits into the left-shifting even-parity vector and right-shifting odd-parity modes. As the mass parameter continuously increases, the greybody factor of the even-parity vector modes exhibits a turning behavior, changing from left-shifting to right-shifting, and then crosses the massless results at a specific set of critical parameters corresponding to $l$, $\mu$, and $\omega$. Beyond these critical values, the behavior of the even-parity modes follows that of the other modes, shifting to higher energy regions compared to the massless vector results. Therefore, a parameter space exists for the Proca particles of even-parity vector modes that includes higher transmission probabilities than those of photons. Note that a similar phenomenon occurs, massive particles having higher transmission probabilities than massless particles within a specific parameter space in a spherically symmetric black hole spacetime, which can also be observed in Dirac perturbations \cite{Cho:2005}, and the turning behavior can also be seen in modified gravity theories as discussed in \cite{BoChNgW2021}.

Furthermore, the correspondence between the greybody factors and the quasinormal modes gained some attention in recent years \cite{Konoplya_2023, Konoplya_2024}, presenting a link between these phenomena through the WKB approximation, despite explicit differences in the boundary conditions. For our results, a comparison with the quasinormal modes in \cite{PreDol2020} shows that the even-parity vector mode exhibits a turning behavior with increasing Proca mass for non-extremal rotating black hole backgrounds, including in the static limit. A direct analysis exploring the relationship between the turning behaviors of the quasinormal modes and the greybody factor is beyond the scope of the current draft and is left for future work. Nevertheless, if these turning behaviors correspond, it may be analogous to the greybody factor of the even-parity vector modes in the Kerr black hole, which also includes a specific parameter space where the transmission probability exceeds that of the massless vector perturbation. This would be of particular interest.

Lastly, since the greybody factor determines the transmission probability of Hawking radiation, it is important to emphasize that the even-parity vector modes can exhibit higher transmission probabilities than those of photons. A natural next step would be to compare the Hawking fluxes of the Proca particle with those of the massless case. Such a comparison could provide further insight and contribute to future experimental efforts aimed at detecting Hawking radiation. For example, this includes the possible detection of Hawking radiation from asteroid-mass primordial black holes \cite{PhysRevLett.126.171101}, the search for echoes or black hole morsels from binary mergers \cite{PhysRevD.108.044047, CACCIAPAGLIA2025117021}, and investigations of the dark photon as a potential dark matter candidate \cite{PhysRevD.101.063030, TANTIRANGSRI2024101432,  PhysRevD.112.023026, Cheung:2025gdn, carenza2025darknessprobinginflationaryera}. 

\section*{Acknowledgements}
SB and CHC are supported in part by the Naresuan University research fund No. R2566C048 and the National Science, Research and Innovation Fund (NSRF) of Thailand, via the Program Management Unit for Human Resources $\&$ Institutional Development, Research and Innovation, grant number B39G670016.
\bibliographystyle{aipnum4-2} 
\bibliography{GB_Proca_Schw}
\end{document}